%% file: ms.tex
\documentclass{aa}

\usepackage{txfonts}

\usepackage{graphicx}	
\usepackage{amsmath}	
\usepackage{amssymb}	
\usepackage{gensymb}
\usepackage{tabulary}
\usepackage{subfigure}
\usepackage{epstopdf}
\usepackage{natbib}
\usepackage{afterpage}
\usepackage{float}
\usepackage{booktabs}
\usepackage{caption}
\usepackage{fancyvrb}
\usepackage{multirow}
\usepackage{hyperref}
\usepackage[capitalise,nameinlink]{cleveref}

\usepackage{xr}		


\defcitealias{cappellari2016structure}{C16}
\defcitealias{graham2018angular}{G18}
\defcitealias{graham2019atechnical}{Paper I}
\defcitealias{graham2019bcatalogue}{Paper II}
\defcitealias{graham2019dclusters}{Paper IV}

\begin{document}


\title{SDSS-IV MaNGA: New benchmark for the connection between stellar angular momentum and environment: a study of about 900 groups/clusters} 
\titlerunning{SDSS-IV MaNGA: 	Stellar angular momentum and environment}

\author{Mark T. Graham$^1$\thanks{E-mail: mark.graham@physics.ox.ac.uk}
\and Michele Cappellari$^1$
\and Matthew A. Bershady$^{2,3}$
\and Niv Drory$^4$}
\authorrunning{M. T. Graham et al.}
\institute{Sub-department of Astrophysics, Department of Physics, University of Oxford, Denys Wilkinson Building, Keble Road, Oxford, OX1 3RH, UK
\and Department of Astronomy, University of Wisconsin-Madison, 475N. Charter St., Madison WI 53703, USA
\and South African Astronomical Observatory, Cape Town, South Africa
\and McDonald Observatory, The University of Texas at Austin, 1 University Station, Austin, TX 78712, USA
}

\date{Accepted XXX. Received YYY; in original form ZZZ}



\abstract
{It has been observed that low redshift early-type galaxies can be separated into slow and fast rotators according to a proxy of specific stellar angular momentum, $\lambda_{R_e}$, which is available only with integral field spectroscopy. There is a clear discontinuity between the two classes which has been attributed to the existence of two distinct formation channels. Detailed studies of a handful of nearby clusters have shown that slow rotators are generally found at the centres of clusters where the number density is highest, whereas the fast rotators trace the trend followed by early-type galaxies of increasing in number with local density.}
{In this paper, we study the environmental distribution of slow and fast rotators using the stellar kinematics of about 3900 galaxies from the Sloan Digital Sky Survey's Mapping Nearby Galaxies at Apache Point Observatory survey.}
{For galaxies in groups closer than $z=0.08$ that are not observed with MaNGA but satisfy the necessary conditions for slow rotators ($M \geq 2 \times 10^{11} \textrm{M}_{\odot}$ and $\epsilon<0.4$), we visually assign slow/fast rotator classifications to obtain a complete sample. Our final catalogue contains about 900 groups of five or more members.}
{For this sample, we observe the kinematic morphology-density relation for each group and find an increasing fraction of massive slow rotators with increasing number density. We provide evidence suggesting that the observed lack of trends in angular momentum with environment at fixed stellar mass intervals is in part due to the fact that the maximum density varies between clusters, and that in fact the locations of massive slow rotators are strongly correlated with the peak densities in galaxy groups and clusters.}
{We conclude that the (projected) number density relative to the cluster peak is more fundamental than the absolute number density in influencing the abundance of slow rotators. We find that the kinematic morphology-density relation does exist at fixed stellar mass, and we rule out the hypothesis that the kinematic morphology-density relation is a result of dynamical friction alone, instead arguing that massive slow rotators grow hierarchically in tandem with their host clusters.}

\keywords{
galaxies: clusters: general --- galaxies: elliptical and lenticular, cD --- galaxies: groups: general --- galaxies: kinematics and dynamics --- galaxies: spiral
}

\maketitle



%

\input{Introduction3}
\input{Methods}
\input{Results}
\input{Comparison}
\input{Discussion}
\input{Conclusions3}

\section*{Acknowledgements}
We acknowledge helpful comments from Kyle Westfall that helped to improve this manuscript. MTG acknowledges support by STFC. MC acknowledges support from a Royal Society University Research Fellowship.\par
Funding for the Sloan Digital Sky Survey IV has been provided by the Alfred P. Sloan Foundation, the U.S. Department of Energy Office of Science, and the Participating Institutions. SDSS acknowledges support and resources from the Center for High-Performance Computing at the University of Utah. The SDSS website is www.sdss.org.\par
SDSS is managed by the Astrophysical Research Consortium for the Participating Institutions of the SDSS Collaboration including the Brazilian Participation Group, the Carnegie Institution for Science, Carnegie Mellon University, the Chilean Participation Group, the French Participation Group, Harvard-Smithsonian Center for Astrophysics, Instituto de Astrofísica de Canarias, The Johns Hopkins University, Kavli Institute for the Physics and Mathematics of the Universe (IPMU) / University of Tokyo, Lawrence Berkeley National Laboratory, Leibniz Institut für Astrophysik Potsdam (AIP), Max-Planck-Institut für Astronomie (MPIA Heidelberg), Max-Planck-Institut für Astrophysik (MPA Garching), Max-Planck-Institut für Extraterrestrische Physik (MPE), National Astronomical Observatories of China, New Mexico State University, New York University, University of Notre Dame, Observatório Nacional / MCTI, The Ohio State University, Pennsylvania State University, Shanghai Astronomical Observatory, United Kingdom Participation Group, Universidad Nacional Autónoma de México, University of Arizona, University of Colorado Boulder, University of Oxford, University of Portsmouth, University of Utah, University of Virginia, University of Washington, University of Wisconsin, Vanderbilt University, and Yale University.\par
This publication makes use of data products from the Two Micron All Sky Survey, which is a joint project of the University of Massachusetts and the Infrared Processing and Analysis Center/California Institute of Technology, funded by the National Aeronautics and Space Administration and the National Science Foundation.





\bibliographystyle{mnras}
\bibliography{MasterBibliography} 





\end{document}

%% file: Introduction3.tex
\section{Introduction}
It has long been understood that the optical morphology of galaxies is related to their local environment \citep{hubble1931relation, zwicky1942clustering, sandage1961hubble, oemler1974clusters, melnick1977morphological}. In a seminal paper, \cite{dressler1980morphology} observed the morphology and position of 6000 galaxies in 55 rich clusters and showed for the first time that there is a fundamental correlation between optical morphology and local projected number density of galaxies. In particular, the spiral population diminishes with increasing local density whereas the fraction of early-type galaxies (ETGs; ellipticals and S0s) increases with increasing local density. This now classic morphology-density relation has been shown to extend to sparser environments \citep{postman1984morphology, helsdon2003density, goto2003density}, and it is found to be already established by $z=1$ \citep{dressler1997density, postman2005density, smith2005density}. The morphology-density relation is believed to be universal, although there is evidence to suggest that the relationship between morphology and clustercentric radius is a more fundamental relation \citep{whitmore1991density, whitmore1993morphological}.\par
Within ETGs, ellipticals are significantly more clustered than S0s \citep{dressler1980morphology}. In fact, the shallow increase in the S0 fraction with increasing local density mirrors the decrease in the spiral fraction suggesting that spirals are transformed into S0s with increasing frequency as the local density increases. Various mechanisms have been proposed to explain this transformation, including ram pressure stripping (e.g. \citealp{gunn1972infall, abadi1999stripping, van2010stripping}), harassment \citep{moore1996harassment}, starvation/strangulation \citep{larson1980disk, dekel2006shock} and tidal interactions (e.g. \citealp{park2007environmental}; see \citealp{boselli2006environmental} for a review on quenching mechanisms of late type galaxies); although it is unclear which mechanisms dominate in groups \citep{smethurst2017quenching} and clusters \citep{taranu2014quenching, boselli2016quenching}. Mergers are also thought to play a role, although galaxy interactions are more likely to take place in groups rather than clusters \citep{barnes2006groups, zabludoff1998groups, hashimoto2000environment, wilman2005groups, martinez2006groups}. There is plenty of evidence to suggest that some S0s are the product of gas-rich (wet) major/minor mergers \citep{barnes2002merging, springel2005merger, robertson2006merger, hopkins2009mergers, moster2011mergers, querejeta2015mergers}, as the gas is able to dissipate kinetic energy while also conserving its mass and specific angular momentum. The process of dissipation encourages disks to form \citep{fall1980formation, abadi1990formation, kormendy1996sequence}, as it tends to reduce the kinetic energy in the direction parallel to the spin axis \citep{cappellari2007sauron}.\par
As a result of the development of integral field spectroscopy (IFS), it has become clear that the S0/elliptical classification of ETGs, which is intended to separate disks and spheroids, is strongly affected by inclination, and in fact two-thirds of all spheroids are actually misclassified face-on, axisymmetric disks \citep{cappellari2011atlas3db}. A more robust classification is based on the specific stellar angular momentum within one effective radius, denoted $\lambda_{R_e}$, which is obtained from observations of the stellar kinematics made possible with IFS \citep{emsellem2007sauron, emsellem2011atlas3d}. A recent revision to the classification scheme defines ETGs as slow rotators if they have low $\lambda_{R_e}$ as a result of highly disordered stellar motions \textit{and} appear round as quantified by the ellipticity within $1R_e$, $\epsilon \equiv 1-b/a$ where $a$ and $b$ are the semi-major and semi-minor axes respectively \citepalias{cappellari2016structure}. All other ETGs (and all spirals) are classified as fast rotators which collectively form a homogenous family of axisymmetric disk galaxies that are supported by rotation and show a range of bulge-to-total ratios (see \citetalias{cappellari2016structure} for a review).\par
In \cite{graham2018angular}, we used a sample of about 2300 galaxies with IFS data of all morphological types to show, without any ambiguity, that fast and slow rotators form two distinct galaxy populations separated by a clean break in angular momentum. They are not a single population within which ETGs transition smoothly from fast to slow rotation. This dichotomy is consistent with previous expectations based on much smaller galaxy samples \citepalias{cappellari2016structure}. Fast and slow rotators are also distinct in stellar mass and the orientation of the spin axis. SRs are generally massive whereas FRs appear at all masses \citep{cappellari2013effect, graham2018angular}. In FRs the spin axis is perpendicular to the disk, but in SRs, the spin axis is randomly orientated \citep{krajnovic2008fast, krajnovic2011atlas3d, fogarty2015sami, graham2018angular} as a result of their triaxiality \citep{weijmans2014atlas, foster2017intrinsic, li2018shape}. The fact that fast and slow rotators form a dichotomy in a number of properties suggests that they have followed separate evolutionary channels.\par
The fact that the fast/slow classification is based on galaxy \textit{structure} prompted \cite{cappellari2011atlas3db} to update the classical morphology-density relation into a \textit{kinematic} morphology-density relation (hereafter kT-$\Sigma$ relation), where ETGs are split according to their angular momentum content. They used the pioneering volume-limited ATLAS$^{\rm{3D}}$ survey of 871 nearby galaxies \citep{cappellari2011atlas3db} of all morphological types to study the kT-$\Sigma$ relation for the first time. The sample included the Virgo cluster, which meant it was possible to compare galaxy populations across over four orders of magnitude in number density, from the field to the dense core of Virgo. They found that the linear \textit{de}crease of spirals mirrors the linear \textit{in}crease in fast rotator ETGs, a result that is consistent with the classical morphology-density relation \citep{dressler1980morphology}. This linear trend breaks inside the core of Virgo (galaxy surface density of $\sim 200 \textrm{ Mpc}^{-2}$), where spirals dramatically decrease in number. \cite{cappellari2011atlas3db} concluded that different processes act in the cluster core to transform spirals into FRs, such as ram-pressure stripping which strips the cold gas as well as prevents cold gas accretion.\par
On the other hand, SRs only appear in any significant fraction in the density bin corresponding to the core of Virgo. \cite{cappellari2011atlas3db} attributed this increase to the high frequency of close encounters and minor mergers that galaxies now lying in the centre of clusters undergo during their evolution, compared to the field. A prerequisite for SR formation is that the merger must be gas-poor i.e. a dry merger \citep{jesseit2009angular, bois2010mergers, bois2011binary, hoffman2010merger, li2017prolate, penoyre2017illustris}. In this scenario, dissipation is not possible as the system is dominated by a collisionless stellar component.\par
The relics of dry mergers can be distinguished from other elliptical galaxies by the presence of a flux deficit in the nuclear region of the surface brightness profile \citep{faber1997galaxies}. This is the reason why the fast/slow rotator dichotomy essentially traces the separation between galaxies with a core and the rest \citep{ferrarese1994elliptical, lauer1995galaxies, lauer2005galaxies, kormendy2009formation} as shown by \cite{lauer2012cores}, \cite{krajnovic2013atlas3d} (page 23) and \cite{krajnovic2019cores}. Cores are thought to arise as a consequence of a merger between two galaxies each containing a supermassive black hole that sinks to the centre of the merger remnant via dynamical friction to form a binary \citep{faber1997galaxies, volonteri2003cores, trujillo2004core}. Successive three-body interactions between stars and the resulting binary eject a stellar mass roughly equivalent to the total mass of the binary \citep{ebisuzaki1991merging, milosavljevic2001formation}. The angular size of these cores is a small fraction compared to the galaxy and are only resolvable with HST imaging \citep{crane1993cores, lauer1995galaxies, faber1997galaxies}. Fortunately, \cite{cappellari2013effect} found that the stellar mass of genuine core SRs exceeds a critical stellar mass of $M_{\rm{crit}}=2\times10^{11} \textrm{ M}_{\odot}=10^{11.3} \textrm{ M}_{\odot}$. Enforcing this criterion allows one to remove ``spurious'' SRs, namely SRs likely not formed through multiple dry mergers, like for example $2\sigma$ galaxies (see page 2 of \citealp{krajnovic2011atlas3d}) in large surveys where HST imaging is not available.\par
Although massive SRs dominate in the centres of clusters, they are not completely absent in the field. \cite{cappellari2013effect} presented the mass-size relation for the ATLAS$^{\rm{3D}}$ field sample and found 10 examples of massive core SRs. \cite{cappellari2013effect} suggested that they form in galaxy groups at high redshifts and sink to the centre of the dark matter halo of the group via dynamical friction \citep{chandrasekhar1943dynamical, white1976dynamical, kashlinsky1986dynamical}. These halos merge with other nearby halos in a hierarchical way \citep{white1978core, white1991galaxy, de2012environmental}. As the massive SRs are essentially at rest within each subhalo and they have a large collisional cross-section, they are able to merge on radial orbits \citep{ramirez1998orbits, boylan2006mergers, von2007cluster}. In contrast, the FRs and spirals have have lower masses, small collisional cross-sections and higher relative velocities, and so are much less likely to merge \citepalias{cappellari2016structure}. Hence, the track on the mass-size relation followed by massive SRs is steeper ($R\propto M$) than for FRs, as they follow different evolutionary paths (\citetalias{cappellari2016structure}, \citealp{graham2018angular}). \cite{scott2014distribution} note that this hierarchical scenario is more realistic than one where SRs are drawn to the centres of clusters by dynamical friction alone. They showed that for a mass-matched sample, SRs are more biased to denser environments compared to FRs. If dynamical friction were indeed the cause, then we would expect SRs and FRs of similar masses to follow the same distribution with respect to local density, given that dynamical friction is only a function of the stellar mass and not the internal kinematic structure of galaxies.\par
Two challenging aspects of studying galaxy trends as a function of environment are that (1) environment has a similar effect on galaxy properties as mass, and (2) many galaxy properties are driven by stellar mass. In fact, the two effects are separable \citep{peng2010galaxy} and the classical morphology-density relation has been observed at fixed stellar mass \citep{bamford2009morphology}. However, the scarcity of massive SRs in a volume limited sample means that a similar study using the kinematic classification scheme requires very large numbers of galaxies. Now, we are in the era of massive IFS surveys. One of these is the SAMI \citep{croom2012sydney,bryant2015sami} survey of 3000 galaxies in the Southern hemisphere. SAMI has a dedicated cluster sample \citep{fogarty2014sami, owers2017cluster, brough2017kinematic} which makes it ideal for studying stellar kinematics of cluster galaxies \citep{fogarty2015sami}. The second IFS survey currently in operation is the Mapping Nearby Galaxies at Apache Point Observatory (MaNGA) survey (\citealp{bundy2015overview}; see \citealp{smee2013multi} for details about the spectrographs, \citealp{drory2015manga} for a complete description of the IFUs, \citealp{law2015observing} for details about the observing strategy and \citealp{yan2016spectrophotometric} for details about the flux calibration) based at the dedicated SDSS telescope \citep{gunn20062} and operating as part of SDSS-IV \citep{blanton2017sloan}. MaNGA aims to observe 10,000 galaxies with IFS across six years \citep{yan2016sdss}. Its unique selection function allows it to observe many more massive galaxies than a volume limited survey, thus making it ideal for studying diverse galaxy populations \citep{wake2017sdss}. Another survey that is relevant to this work is the MASSIVE survey \citep{ma2014massive} which targets very massive galaxies ($M \gtrsim 10^{11.5} \textrm{ M}_{\odot}$) for a volume limited sample and is designed to complement the ATLAS$^{\rm{3D}}$ survey ($M \lesssim 10^{11.5} \textrm{ M}_{\odot}$).\par
Three separate groups have studied the kT-$\Sigma$ relation at fixed stellar mass using data from the MaNGA \citep{greene2017kinematic}, SAMI \citep{brough2017kinematic} and MASSIVE \citep{veale2017angular} surveys and they have found very little or no correlation between a galaxy's angular momentum and the local galaxy density. These results have been corroborated by \cite{choi2018spin}, who separated galaxies in the Horizon-AGN simulation by cluster, groups and field and found very little trend in $\lambda_{R_e}$ with the number density at fixed stellar mass. These results have been interpreted as evidence that the kT-$\Sigma$ relation is in fact driven by segregation of stellar mass, whereby the most massive galaxies sink to the centre of the potential well of galaxy groups and clusters and so find themselves at the highest densities \textit{as a result} of their stellar mass \citep{brough2017kinematic}. In actual fact, there is no clear reason why such a trend should exist, because by only considering galaxies in a certain mass bin, the resulting subsample is naturally restricted in terms of the environment. One possible interpretation of these results is that dense environments are not required to form slow rotators. This may seem to be in tension with the results from nearby clusters, which clearly illustrates that massive core SRs have a distinctly different spatial distribution to FRs and spirals as a result of their evolution \citep{cappellari2013effect, d2013fast, houghton2013densest, scott2014distribution, cappellari2016structure}. However, these two different explanations for the distribution of massive SRs are not necessarily inconsistent, as they merely represent two alternative ways of assessing the environmental dependence of galaxy angular momentum (i.e. whether or not to fix the stellar mass).\par
It should be noted that \cite{lee2018environment} studied the relationship between galaxy spin and small- and large-scale environment for a sample of $\sim1500-1800$ MaNGA galaxies. They found that galaxy spin is insensitive to the large-scale environment as probed by a background mass density of 20 nearby galaxies. However, they found some evidence that $\lambda_{R_e}$ for late-type galaxies is sensitive to the distance to the nearest neighbour (small scale environment), if the neighbours are early-type galaxies. They suggest that their results indicate that hydrodynamical interactions can play a part in reducing the spin of regular rotators. However, they make no conclusions about the distribution of fast and slow rotators with respect to their environment. Moreover, \cite{bernardi2019environment} studied the relationship between environment and angular momentum for a sample of massive ($\log(M) \gtrsim 11$) FR and SR ETGs but were unable to draw firm conclusions due to small number statistics in each mass-environment bin.\par
Up until now, studies have fallen into two camps: detailed studies of a few individual clusters with complete kinematic observations, including the Fornax, Coma and Virgo clusters, or extensive studies with the MaNGA, SAMI and MASSIVE surveys of many clusters each with incomplete kinematic classifications (although SAMI does have a cluster sample). In this paper, we try to combine the strengths of both approaches by studying many clusters, each with complete kinematic classifications in order to gain valuable insight into the role of environment in galaxy transformation and kinematic structure.\par
We base our analysis on stellar kinematics of about 4500 galaxies released as part of the MaNGA Product Launch 7 (MPL-7) internal data release. MaNGA is ``agnostic'' to environment (see \citealp{smethurst2018quenching}), meaning that any selection in environment, if present, arises indirectly as a result of the other selection criteria (for example stellar mass). MaNGA does not target specific clusters, apart from one ancillary programme \citep{gu2018coma}, and so galaxies are selected from multiple distinct groups. Hence, we do not have complete stellar kinematic observations for any clusters which play host to MaNGA galaxies. To overcome this, we have developed a novel method to assign angular momentum classifications to all galaxies for which we do not have stellar kinematics, which we describe in detail in \cite{graham2019bcatalogue} (Paper II). As a result, we have a sample of about 14,000 galaxies in groups of $z \leq 0.08$ of which \textit{$\sim98.8\%$} have accurate angular momentum classifications. These are galaxies which either (1) have stellar kinematic observations, (2) do not satisfy the criteria for genuine SRs ($\log(M)<11.3$ or $\epsilon \geq 0.4$) and so by definition must be FRs, or (3) have been visually classified as FRs with 94\% accuracy. The remaining $\sim1.2\%$ (about 200 galaxies) satisfy the criteria for genuine SRs ($\log(M)\geq11.3$ and $\epsilon < 0.4$) \textit{and} have been visually classified as SRs. However, for this specific case, the classification is only accurate about half the time. This is the largest sample of its kind and will provide the basis for the in-depth study of angular momentum and environment that we present here.\par
In \cite{graham2019dclusters} (Paper IV), we present schematics for ten clusters in our sample, highlighting the locations of the massive SRs in particular. In \cref{sec:results}, we study the environmental properties of galaxies for all group sizes taken from our catalogue. We also compare with previous works by attempting to reproduce their results using similar methodologies (\cref{sec:comparisonwork}).\par

%% file: Methods.tex
\section{Methodology}
The results presented in this paper rest on several methodologies and conclusions outlined in the two previous papers. We now summarise the content of these papers while preserving the logical order of the key steps.

\subsection{Companion works}
\label{sec:companion}
\begin{enumerate}
\item In sec. 2 of \citetalias{graham2019bcatalogue}, we addressed the incompleteness in the NSA as a function of redshift by combining the NSA with the SDSS photometric catalogue. Upon visual inspection of the photometric catalogue, we realised that the automatic imaging pipeline does not offer complete accuracy even when the recommended cleaning criteria of SDSS are applied. To improve the sample of objects, we proposed empirical criteria based on photometric criteria including five SDSS colours, the error in photometric redshifts and the Petrosian radius in the $r$-band, $R_{\rm{r,Petro}}$. We demonstrated that the empirical criteria are independent of the SDSS photometric flags and only target genuinely suspicious objects, leaving the vast majority of genuine galaxies alone.
\item As galaxies which are not in the NSA do not have stellar mass estimates, we used dynamical masses from MaNGA to fit a stellar mass-luminosity relation using the absolute $r$-band magnitude $M_r$ (\citetalias{graham2019bcatalogue}). (Previously this had been done with $M_{K_s}$ from the Two-Micron All Sky Survey (2MASS) Extended Source Catalogue (XSC) but not all objects in the SDSS photometric catalogue have 2MASS photometry.)
\item For each MaNGA galaxy, we define a cylinder with radius 10 comoving Mpc and height $h=2\Delta V$ km s$^{-1}$ centred on the MaNGA galaxy, where $\Delta V$ is specified between 300 and 3000 km s$^{-1}$ based on the empirical position-velocity distribution of the local \textit{spectroscopic} galaxies from the NSA. Photometric galaxies can have redshift errors orders of magnitude larger than the height of the cylinder and so photometric interlopers are highly likely.
\item To improve the selection by removing potential interlopers, we make the physically motivated assumption that all neighbours must have a luminosity $M_r\leq-18$ (corresponding to $M \sim 7.4\times10^{9} \textrm{ M}_{\odot}$) \textit{assuming they are at the redshift of the MaNGA galaxy}. Hence, background interlopers with a given intrinsic luminosity at their most likely redshift will have a \textit{lower} intrinsic luminosity when they are ``brought forward'' to the redshift of the MaNGA galaxy. We ignore the velocity cylinder in this step, although a galaxy could potentially satisfy the luminosity cut at $V>0$ but not $V=0$ where $V$ is the velocity with respect to the MaNGA galaxy. However, the chances of this happening are low and in any case, the error in magnitude is likely to be larger than the difference in the apparent magnitude of the luminosity cut between $V=0$ and $V=\Delta V$.
\item We run the local group finder \texttt{TD-ENCLOSER} presented in \citetalias{graham2019atechnical} on the surviving galaxies i.e. galaxies that lie within the velocity cylinder and satisfy the luminosity cut. We treat all surviving galaxies the same regardless of the uncertainty on the redshift. We identify the \textit{set of neighbours} that encloses the ``host'' MaNGA galaxy i.e. the galaxy at the centre of the velocity cylinder. We define this galaxy to be the ``host'' galaxy of its set because its redshift defines the \textit{apparent} magnitude of the luminosity cut as well as the position of the velocity cylinder. We use this terminology primarily to differentiate this galaxy from other MaNGA galaxies that may lie in the same set.
\item In \citetalias{graham2019bcatalogue}, we describe how we construct the group catalogue which we use in this paper and introduce the environmental parameters that we provide in the catalogue. These are:
\begin{itemize}
\item $\mathcal{N}$: Group richness, or number of members in a group.
\item $\Sigma_3$ ($\Sigma_{10}$): Projected number density within a circle centred on a galaxy with a radius equal to the distance to the third (tenth) nearest neighbour.
\item $\Sigma_3^{\rm{rel}}$: Relative projected number density compared to the peak $\Sigma_3$ within a group.
\item $D_{\rm{cen}}$: Projected distance from a galaxy to the central galaxy of its enclosing group. The central galaxy is defined as the galaxy closest to the peak of the density field, and is unrelated to its mass (for example, it is not a brightest cluster galaxy).
\item $D_{\rm{cen}}^{\rm{norm}}$: Projected distance from a galaxy to the central galaxy of its enclosing group normalised by the 90th percentile of the group.
\end{itemize} In \citetalias{graham2019bcatalogue}, we define a redshift cut of $z=0.08$ beyond which we exclude galaxy groups from our science sample. Our primary motivation for introducing a cut is that the accuracy of the groups decreases substantially as redshift increases due to the incompleteness of the NSA. Groups above this redshift will be dominated by galaxies with photometric redshifts and so the confidence in any conclusions drawn from a sample that includes these groups will be affected. We acknowledge that the majority ($\sim 57 \%$) of confirmed SRs in MaNGA are beyond this redshift as a result of the MaNGA selection function, but the SR sample at $z \leq 0.08$ is still large enough to allow us to draw statistical conclusions.
\item A second motivation for limiting our final sample below $z=0.08$ is that we visually classify slow/fast rotator candidates for the subset of galaxies that \textit{could} be genuine SRs (i.e. $\log(M)\geq 11.3$ and $\epsilon<0.4$). We base our classification on the SDSS photometry, the quality of which depends on redshift, by looking for certain morphological features such as an extended stellar halo to identify SR candidates. As a test, we guess the classifications for massive, round MaNGA ETGs for which we have the correct answer, and we find that given a guess of FR, we are correct in 92\% of cases. This is crucial for our conclusions as it suggests that we can rule out galaxies that are definitely \textit{not} SRs with almost complete accuracy. In contrast, given a guess of SR, we only achieve 48\% accuracy meaning that our misclassification is essentially random: half of all galaxies we identify as SRs will be confirmed as FRs if stellar kinematics become available. We only identify SR/FR candidates for galaxies which are found in our groups i.e we do not classify the surviving galaxies for (iv) that do not lie in any of our groups.
\item In preparation for the science analysis in this section, we identify sets which may be duplicates of the same \textit{intrinsic groups}. In \citetalias{graham2019atechnical}, we present a simple algorithm which finds all possible duplicates (or even sets which are linked by a thread of MaNGA galaxies) and selects one or more representative groups using a combination of group richness and a random selector. We thus obtain a group catalogue where each MaNGA galaxy appears only once, although there will be many galaxies which are not in MaNGA that appear in multiple groups.
\end{enumerate}

%% file: Results.tex
\section{Results}
\label{sec:results}
We start by updating the $(\lambda_{R_e}, \epsilon)$ diagram for about 4000 galaxies in MPL-7. This sample is nearly twice as large than the one we presented in \citetalias{graham2018angular}. We then present the results from our complete kinematic study.

\subsection{Stellar angular momentum of about 4000 galaxies}
We update the right hand side of fig. 5 in \citetalias{graham2018angular} for a sample of 4003 galaxies which we present in \cref{fig:lambda_ellip}. We apply the correction introduced in appendix C of \citetalias{graham2018angular} to account for the effect on $\lambda_{R_e}$ due to the atmospheric seeing. This correction was tested by \cite{harborne2019spin} on realistic galaxy models with a range of bulge-to-total mass ratios and Sersic indices and which are completely independent of the JAM models used to derive the correction. They generated 1425 simulated observations from their models according to the SAMI specifications and measured $\lambda_{R_e}$ for a range of seeing conditions. They found that the atmospheric correction does a good job in recovering the intrinsic $\lambda_{R_e}$ with no systematic deviations for a range of PSF widths (see their fig. 10). They also confirmed the level of scatter seen in fig. C4 of \citetalias{graham2018angular} due to the inclination.\par
The upper envelope does not extend as high in \cref{fig:lambda_ellip} compared to the right hand side of fig. 5 in \citetalias{graham2018angular}. In short, the reason for the discrepancy is because here we use a fixed value for the correction required to obtain the astrophysical velocity dispersion, $\sigma_{\rm{corr}}$ of 40 km s$^{-1}$, whereas in \citetalias{graham2018angular}, we use the value provided by the DAP, which has a median value of $\sigma_{\rm{corr}}=51 \textrm{km s}^{-1}$ for MPL-5. We do not use $\sigma_{\rm{corr}}$ provided for MPL-7 for the reasons outlined below.\par 
The DAP calculates $\sigma_{\rm{corr}}$ as the average quadrature difference between the resolution vectors of the MILES templates and the MaNGA spectra (see eq. 7 of \cite{westfall2019pipeline}). Briefly, the DAP fits \texttt{pPXF} \citep{cappellari2004parametric, cappellari2017improving} to the MILES template and the galaxy spectra obtaining $\sigma$ for both. The quadrature difference between $R^{-2}_{\rm{MaNGA}}$ and $R^{-2}_{\rm{template}}$ where $R = \lambda/\Delta \lambda$ is calculated for each wavelength channel. All quadrature differences are summed and the sum is multiplied by a factor proportional to the number of wavelength channels fitted by \texttt{pPXF}. In this way, the average instrumental resolution is taken across all wavelength channels. This method was tested on galaxy models and real galaxies and was found to be robust (Westfall, private communication). In MPL-5, the instrumental resolution vector was taken to be a single number for MaNGA ($\sigma_{\rm{inst}}^{\rm{MaNGA}} \sim 72 \textrm{ km s}^{-1}$; \citealp{law2016data}), and as $\sigma_{\rm{inst}}^{\rm{MILES}} \sim 50 \textrm{ km s}^{-1}$ \citep{sanchez2006library, falcon2011updated} for MILES, the MPL-5 correction took a value of $\sigma_{\rm{corr}}^{\rm{MPL-5}}\approx 52 \textrm{ km s} ^{-1}$.\par
Between MPL-5 and MPL-7, the resolution vector used by the DAP for the calculation of the correction changed in order to make the calculation more accurate. However, the median for the MPL-7 correction, $\sigma_{\rm{corr}}^{\rm{MPL-7}}$, is $\sim33\textrm{ km s}^{-1}$ which is significantly smaller than the MPL-5 correction (the correction was expected to change only by a few km s$^{-1}$). If we use $\sigma_{\rm{corr}}^{\rm{MPL-7}}$, then we find that the FRs do not fill the upper envelope as completely as is shown in \autoref{fig:lambda_ellip}, only reaching $\lambda_{R_e} \approx 0.85$ (excluding a handful of outliers). Although we know that $\sigma_{\rm{corr}}^{\rm{MPL-5}}$ is also not exactly equal to the true velocity dispersion correction (see Appendix B of \citealp{graham2018angular}), we also know that $\sigma_{\rm{corr}}^{\rm{MPL-7}}$ is not equal to the truth either. If we assume that the expected instrumental resolution of MPL-7 is 65 km s$^{-1}$ (Belfiore, private communication), then the expected correction for MPL-7 is $\sigma_{\rm{corr}}^{\rm{MPL-7}} = \sqrt{65^2-50^2}\sim 40 \textrm{ km s}^{-1}$. This value was indeed found by comparing the DAP derivation of the correction to a ``direct" calculation of the correction obtained by fitting the MILES spectra to the MaNGA stellar library (MaStar; \citealp{yan2018library}) spectra (fig. 17 of \citealp{westfall2019pipeline}). Hence, we choose a fixed correction of 40 km s$^{-1}$ which is somewhere in between the MPL-5 and MPL-7 correction.\par
By using a correction that is smaller than the one from MPL-5, we can neatly explain why we observe a lower FR envelope in \cref{fig:lambda_ellip} compared to the right hand side of fig. 5 in \citetalias{graham2018angular}. We can also explain why we observe a peculiar gap between the magenta line and the isotropic rotator line (green) in fig. 5 of \citetalias{graham2018angular}. In \citetalias{graham2018angular}, we attribute this gap to the possible circularisation of the ellipticity due to seeing or perhaps the larger distances in MaNGA compared to ATLAS$^{\rm{3D}}$ and CALIFA, which have better quality data. We now understand that $\lambda_{R_e}$ was overestimated due to an overestimation of $\sigma_{\rm{corr}}$ which was brought to our attention with the development of the new, more robust correction. More work has been done since the release of MPL-7 to understand this complex issue. Regardless, these effects do not affect the results presented in \cite{graham2018angular} regarding the detection of the bimodality, as galaxies near the SR/FR boundary do not have low velocity dispersion.\par
We only plot 4003 galaxies in \cref{fig:lambda_ellip}, but there are 4597 galaxies in MPL-7. About 600 galaxies do not satisfy the quality control criteria given in sec. 3.6 of \citetalias{graham2018angular} and so their $\lambda_{R_e}$ and/or $\epsilon$ are not deemed to be accurate enough to place on the $(\lambda_{R_e},\epsilon)$ diagram. Many of these galaxies are mergers or have flagged kinematics for example. For these galaxies, we classify them as fast or slow rotators using the method from \cite{graham2019bcatalogue} as well as the stellar kinematics. Because we do have the stellar kinematics for these galaxies, we do not consider them to be affected by the misclassification bias because we have extra information from the kinematics.

\begin{figure}
\centering
\includegraphics[width=0.49\textwidth]{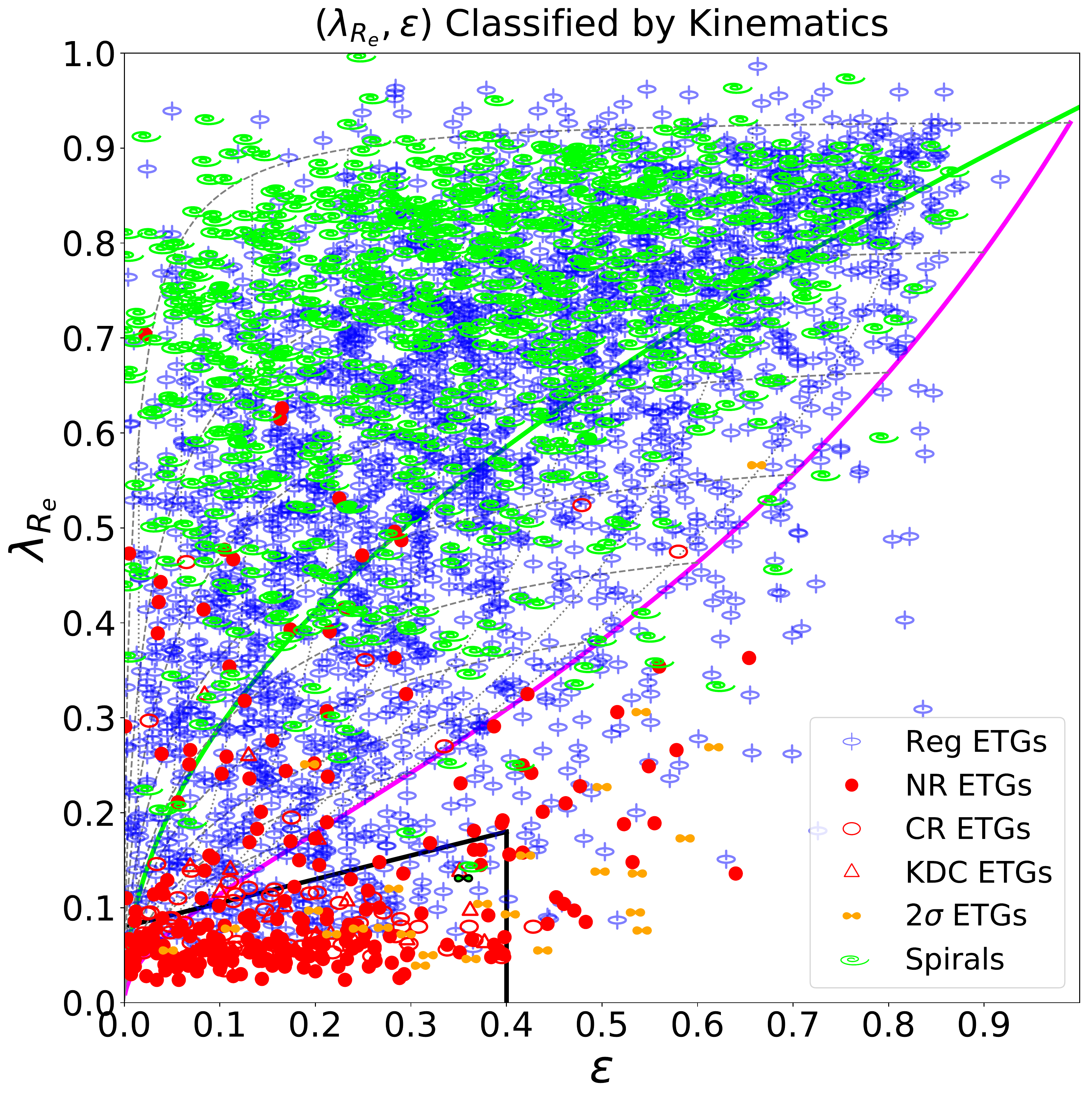}
\caption[$(\lambda_{R_e},\epsilon)$ diagram for about 4000 galaxies]{\textbf{$(\lambda_{R_e},\epsilon)$ diagram for about 4000 galaxies.} We plot the proxy for specific stellar angular momentum $\lambda_{R_e}$ and the ellipticity $\epsilon$ for 4003 galaxies in MPL-7. The lines and symbols are the same as in fig. 5 of \citetalias{graham2018angular}. All points outside the SR region have been corrected for atmospheric seeing using eq. 5 of \citetalias{graham2018angular}.}
\label{fig:lambda_ellip}
\end{figure}

\subsection{Results from a volume-limited sample of about 14000 galaxies}
\label{sec:results_comp}
The MaNGA survey selection is agnostic to environment, meaning that while the survey has no selection in environment, the flat selection in stellar mass means that the number of dense environments will be higher than for a volume limited sample. There is nothing special about the groups in our catalogue, or in other words, there is no reason why the galaxy groups contained in this sample should be different from any other groups which do not enclose any MaNGA galaxies (and therefore are not presented).\par

\subsubsection{F(SR) vs $\log(\Sigma_3)$}
\begin{figure*}
\centering
\includegraphics[width=0.95\textwidth]{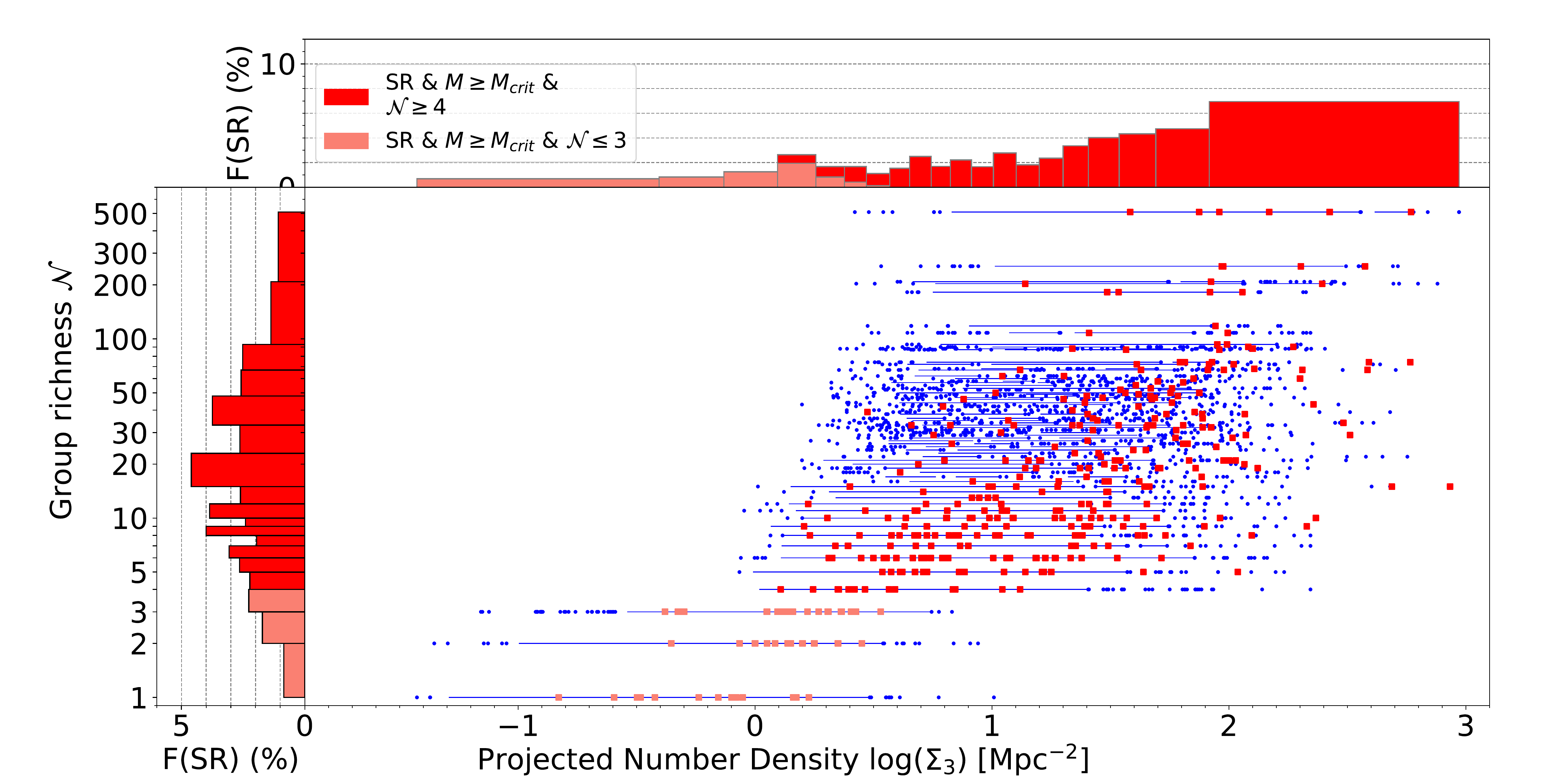}
\caption[Kinematic morphology-density relation]{\textbf{Kinematic morphology-density relation.} We plot group richness $\mathcal{N}$ as a function of $\log{\Sigma_3}$ for all galaxies in our sample. The $y$-axis is shown as a logarithmic scale for visual purposes. Each horizontal line corresponds to a single group richness. In the lower half of the diagram, there will be multiple groups on a single line, whereas for the upper half, there will be between one to a few groups on a single line. For $1 \leq \mathcal{N} \leq 3$, $\log(\Sigma_3)$ necessarily includes galaxies which lie in the field or in nearby groups. Massive SRs in these groups are coloured salmon, while massive SRs in groups of $\mathcal{N} \geq 4$ are coloured red. FRs are coloured blue and in places where the density of blue points is too high, we instead show a thin line. The medians for the blue and red/salmon population for $\mathcal{N}$ and $\log(\Sigma_3)$ are shown as blue and red dashed lines respectively. We do not include isolated galaxies for the horizontal median as there are roughly equal numbers of blue points just on the $\mathcal{N} = 1$ line compared to $\mathcal{N} > 1$. We also plot the F(SR) histograms along both axes, with the salmon population separated from the red population.}
\label{fig:Group_mem_sigma_3}
\end{figure*}

In \cref{fig:Group_mem_sigma_3}, we present $\mathcal{N}$ vs $\log(\Sigma_3)$ for 14093 galaxies. Each point corresponds to a galaxy, but in places where the density of blue points is too high, we simply draw a line. All the points on the same horizontal line do not necessarily belong to the same group; they simply share the same group richness $\mathcal{N}$. Galaxies are coloured red or salmon if they are SRs (MaNGA = \citetalias{cappellari2016structure} box criterion, non-MaNGA = visual classification and $\epsilon < 0.4$) \textit{and} have a stellar mass greater than $M_{\rm{crit}}$. We separate SRs that lie in groups of four or larger (red) from those that are either isolated or lie in pairs or triplets (salmon). This is purely because $\Sigma_3$ includes galaxies from nearby groups when $\mathcal{N} \leq 3$, and so there is a large shift in the $\log(\Sigma_3)$ distribution between $\mathcal{N}=3$ and $\mathcal{N}=4$. (Later on we change the separation for a different reason.) All other galaxies are coloured blue.\par

First, we consider the $x$-axis histogram in \cref{fig:Group_mem_sigma_3}, namely the kT-$\Sigma$ relation for massive SRs. In all cases, F(SR) is defined as the fraction of massive SRs compared to all galaxies i.e. $F(SR)=N(SR)/[N(SR)+N(FR)]$ where N(FR) is the number of galaxies which are \textit{not} massive SRs i.e. are FRs or are less massive than $M_{\rm{crit}}$ but are classified as a SR with kinematics. In general, F(SR) increases from $\sim1\%$ at low number density to $\sim7\%$ at high number density. We are close to the value of 4\% reported by \cite{cappellari2011atlas3db} for F(SR) in the field (compared to all galaxies). For ETGs only, \cite{houghton2013densest} and \cite{cappellari2011atlas3db} find a fraction of $15\%$ if the fraction is defined as N(SR)/N(FR) (where FR does not include spirals). We do not have access to Hubble morphologies for all galaxies in our sample, and so cannot confirm this fraction. However, as we are consistent with \cite{cappellari2011atlas3db} for all galaxies (including spirals), we have every reason to agree with the 15\% fraction for ETGs in the field.\par
However, we can confidently rule out a fraction of 15\% for $2 \lesssim \log(\Sigma_3) \lesssim 3$ as was claimed by \cite{houghton2013densest}. We can do this because spirals are nearly absent in this regime and so we are essentially only considering ETGs \citep{cappellari2011atlas3db}. We also have very good number statistics with about 700 galaxies in each $\Sigma_3$ bin. In \citetalias{graham2019dclusters}, we present schematics for the most massive clusters in our sample. In some cases, we combine clusters which are geographically adjacent, but in \cref{fig:Group_mem_sigma_3}, we keep them separate. The largest cluster in our sample is the Coma cluster, which occupies the top line at $\mathcal{N}\approx 500$ in \cref{fig:Group_mem_sigma_3}. There are six massive SRs, only one of which is a candidate. The total F(SR) for all galaxies in Coma is 1\%, while for the highest density bin, F(SR) is approximately 4/200 = 2\%. The difference between this fraction and the 15\% fraction quoted by \cite{houghton2013densest} for all densities cannot be attributed to the fact that \cite{houghton2013densest} considered SRs of all masses, since the number of (non dry merger relic) SRs below $M_{\rm{crit}}$ is small. Hence, we can confidently rule out a constant fraction F(SR) of 15\% \textit{in clusters}.\par

\begin{figure*}
\centering
\includegraphics[width=0.95\textwidth]{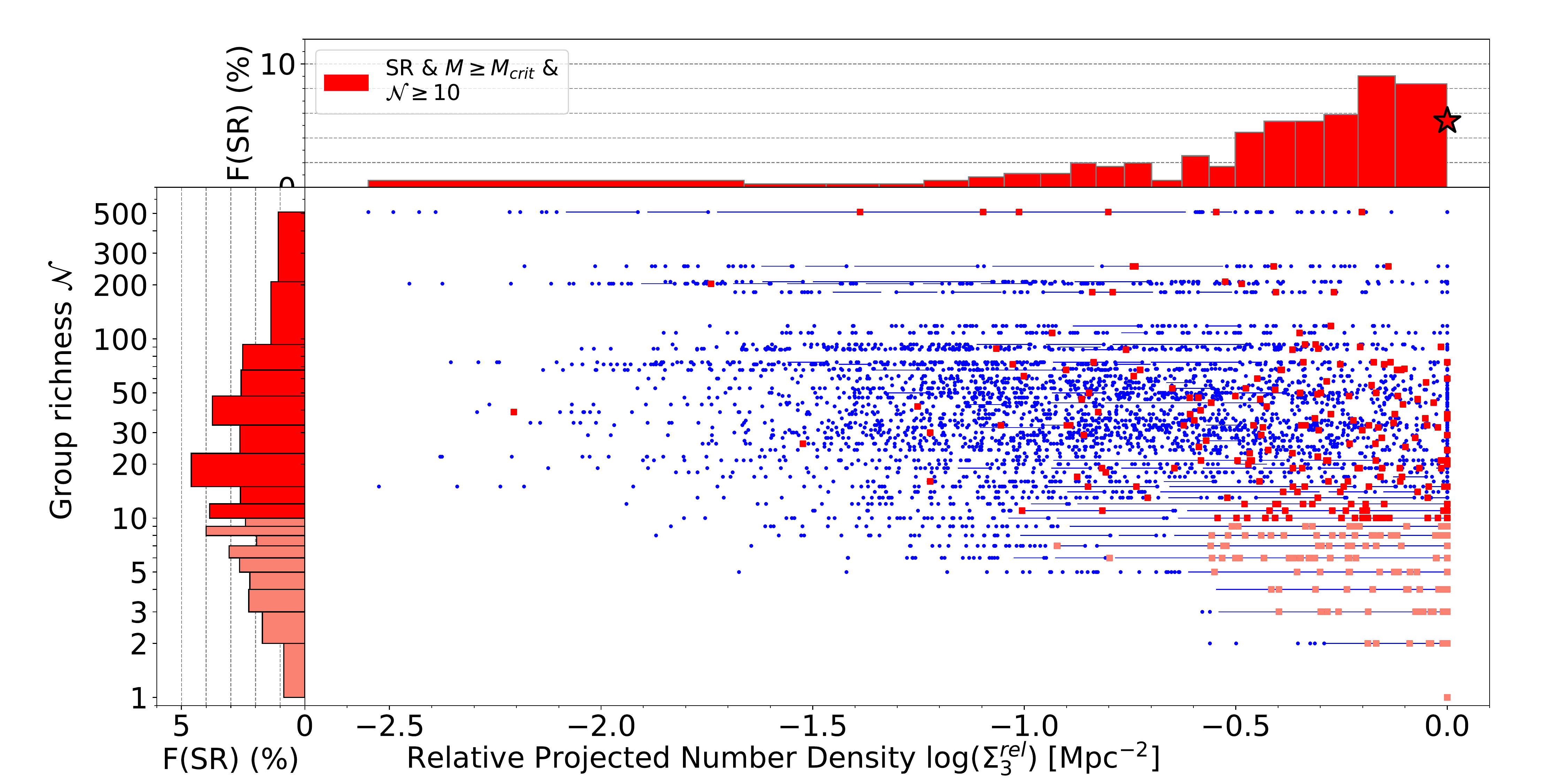}
\caption[Group richness vs relative number density]{\textbf{Relative number density.} The same as \cref{fig:Group_mem_sigma_3} except that the $x$-axis now shows the relative number density $\log(\Sigma_3^{\rm{rel}})$. Unlike \cref{fig:Group_mem_sigma_3}, groups with $\mathcal{N} < 10$ are distinguished from $\mathcal{N} \geq 10$. The histogram along the $x$-axis only considers groups with $\mathcal{N} \geq 10$ to avoid contamination from small groups and isolated galaxies. The red star shows the fraction for galaxies with $\log(\Sigma_3^{\rm{rel}})=0$ and $\mathcal{N} \geq 10$.}
\label{fig:Group_mem_sigma_3_rel}
\end{figure*}

There is a small bump at $\log(\Sigma_3)\approx0.3$ corresponding to the number density where the $\mathcal{N} \leq 3$ and $\mathcal{N} \geq 4$ overlap. Indeed, the two coloured histograms should be considered separately. There is a very weak trend for $\mathcal{N} \leq 3$ which implies that F(SR) for isolated galaxies, galaxy pairs or triplets somehow depends on the distance to the nearest galaxy or group. Assigning galaxies to groups is not a solved problem. Our group-finder algorithm, described in \citetalias{graham2019atechnical}, clips outliers from the edge of groups which then go on to form new, smaller groups. However, the number density is unchanged and so we find a few small groups which are close to nearby, denser groups. This artificially causes the slight bump at $\log(\Sigma_3)\approx0.3$.\par

\subsubsection{F(SR) vs $\mathcal{N}$}
We plot the fraction of massive SRs as a function of group richness $\mathcal{N}$ as a histogram along the $y$-axis in \cref{fig:Group_mem_sigma_3}. The fraction increases from 1\% for isolated galaxies (comparable to the 2\% reported by \citealp{cappellari2011atlas3db}) to $\sim6\%$ for $\mathcal{N} \approx 10 - 30$. Beyond $\mathcal{N}=30$, the fraction decreases to about $1-2\%$ at $\mathcal{N} \approx 200 - 500$. Each bin contains at least 500 galaxies and so the measurement is of high accuracy. The fact that a peak exists at intermediate $\mathcal{N}$ suggests that there is an optimal group size where the process which facilitates the formation of SRs is at its most efficient. 
There is a chance that the height of this peak is more pronounced by virtue of the fact that our groups are not selected at random, but are only in our study because they contain MaNGA galaxies. Since the MaNGA galaxies are flat in stellar mass, we may have more groups with SRs at a given $\mathcal{N}$ than for a volume limited sample. However, as already discussed, the groups themselves are not special in any way and so we are confident that the peak is real. 

\subsubsection{F(SR) vs $\log(\Sigma_3^{\rm{rel}})$}
As discussed in sec. 3.4.1 of \citetalias{graham2019bcatalogue}, the fraction of SRs at a given $\log(\Sigma_3)$ is not a constant, but can vary by up to about $33\%$. We defined a new parameter $\Sigma_3^{\rm{rel}}$ which shifts all clusters to a common baseline such that the galaxy at the peak density has a value of $\log(\Sigma_3^{\rm{rel}})=0$ Mpc$^{-2}$. Clearly, this does not make sense for individual galaxies or galaxies in small groups, and so we only calculate F(SR) as a function of $\log(\Sigma_3^{\rm{rel}})$ for groups with 10 or more members as indicated in \cref{fig:Group_mem_sigma_3_rel}. There is a clear difference between F(SR) as a function of $\log(\Sigma_3^{\rm{rel}})$ compared to $\log(\Sigma_3)$. For the seven bins where $\log(\Sigma_3^{\rm{rel}}) \gtrsim -0.5$ Mpc$^{-2}$, F(SR) fluctuates between 5\% and 10\%. For all other bins, F(SR) is less than 3\% and for $\log(\Sigma_3^{\rm{rel}}) \lesssim -1$ Mpc$^{-2}$, the fraction is less than about 1\%. The striking variability in F(SR) above and below $\Sigma_3^{\rm{rel}} \sim -0.5$ Mpc$^{-2}$ reflects what we see by eye (see \citetalias{graham2019dclusters}). At the locations of peak density in groups and clusters, the fraction of galaxies that are massive SRs is a factor of 5-10 higher than in regions which are underdense compared to the peak.\par
The red histogram ($\mathcal{N} \geq 4$) steadily increases from $\log(\Sigma_3)\approx0.1$ up to the maximum number density, in agreement with previous results. The kT-$\Sigma$ relation presented here is an upper limit, as some of the galaxies we identify as SRs will inevitably be misclassified. However, another source of uncertainty to consider is that due to the errors in ellipticity and $\lambda_{R_e}$, although the latter only applies to galaxies with $\lambda_{R_e}$ measurements. To simulate the effect due to random errors, we randomly perturb $\epsilon$ values for all galaxies 100 times within an error $\Delta \epsilon$, which can be calculated from fig. 2 of \citetalias{graham2018angular} as $2.6 \times 0.056 / \sqrt{2}\sim0.1$. For MaNGA galaxies only, we randomly perturb $\lambda_{R_e}$ within $\Delta \lambda_{R_e}$. For each dataset, we select all galaxies that satisfy the SR criteria. Galaxies which lie close to the border of the SR region risk becoming FRs and vice versa. The uncertainty in $\lambda_{R_e}$ due to random noise in $V,\sigma$ etc. is illustrated in fig. 29 of \cite{yan2016sdss} where $\Delta \lambda_{R_e}$ is shown as a function of $\lambda_{R_e}$. We fit the $2\sigma$ percentiles by eye to be $\Delta \lambda_{R_e}=0.02 + 0.08\lambda_{R_e}$. Clearly this breaks down at $\lambda_{R_e}\sim1$ as random errors will not push $\lambda_{R_e}$ above 1. However, this is only a crude fit that needs only to be realistic for low $\lambda_{R_e}$ values near the SR regime. We add an extra source of error due to the beam smearing correction (see sec. 3.8 of \citetalias{graham2018angular}). The correction gives an upper limit to the true value of $\lambda_{R_e}$. We find that the mean error in the $-\lambda_{R_e}$ direction for MPL-5 is 0.03 (see table 2. of \citetalias{graham2018angular}). Hence, our random error on $\lambda_{R_e}$ is $[+(0.02 + 0.08\lambda_{R_e}),-(0.05 + 0.08\lambda_{R_e})]$.\par

\begin{figure*}
\centering
\includegraphics[width=0.95\textwidth]{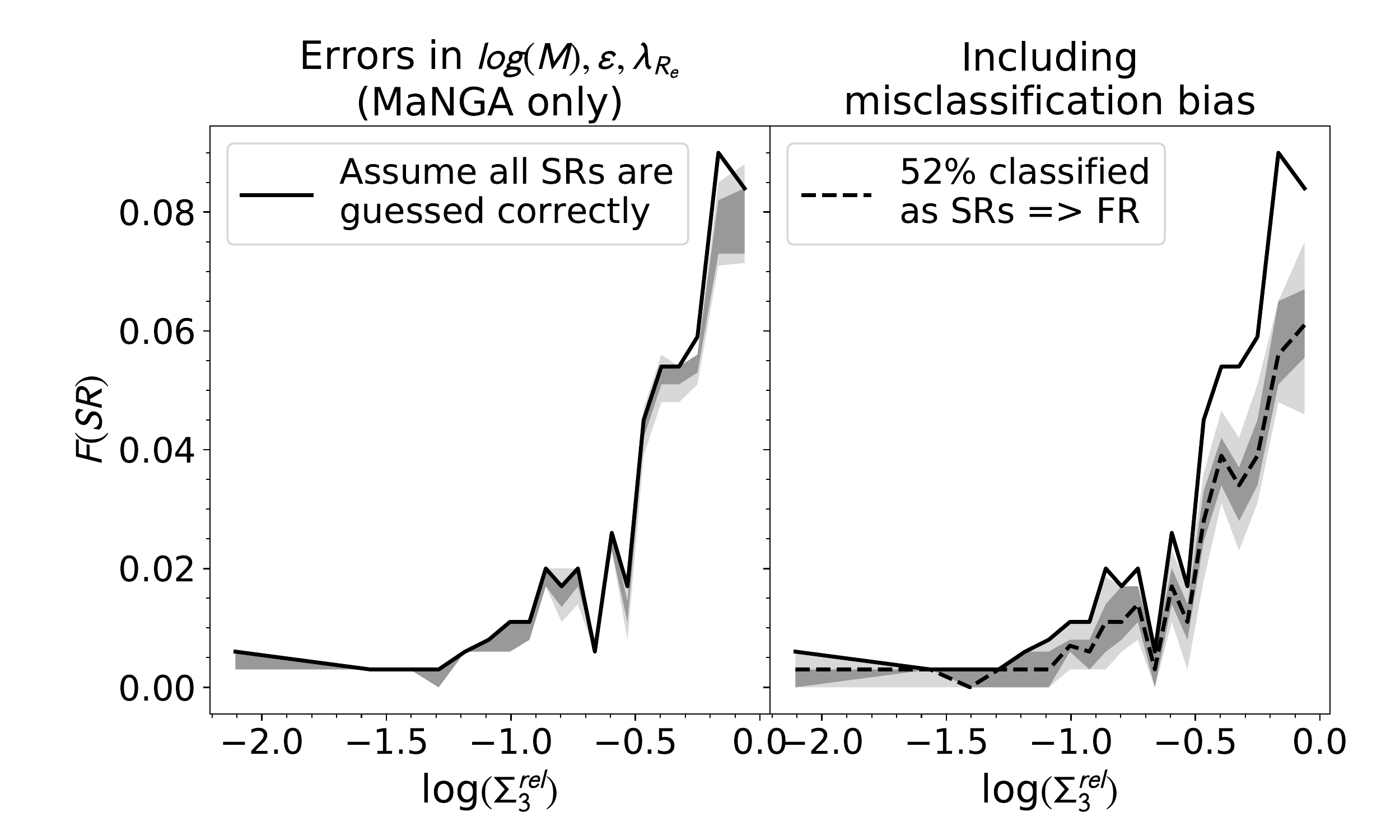}
\caption[Uncertainty in the kT-$\Sigma$ relation]{\textbf{Uncertainty in the kT-$\Sigma$ relation.} \textit{Left}: In black, we plot the same relation seen in the upper histogram of \cref{fig:Group_mem_sigma_3}. In dark and light grey, we plot the inner 68th and 95th percentiles for the fractions in each bin from 100 perturbed samples (see the text). \textit{Left}: The black line is the same as left, and the black dotted line is the 50th percentile now assuming that 48\% of the SR \textit{candidates} are misclassified and are therefore FRs. Again we show the 68th and 95th inner percentiles. Here we also include the errors shown on the left hand side.}
\label{fig:random_sr}
\end{figure*}

In the left hand side panel of \cref{fig:random_sr}, we show the uncertainty in the kT-$\Sigma_3^{\rm{rel}}$ relation due to random measurement errors. For galaxies not in MaNGA, we only consider errors in ellipticity, assuming that the stellar mass and all visual classifications are correct. (The way to simulate the errors in $\lambda_{R_e}$ for these galaxies would be to classify them as FRs or SRs with 100 independent classifiers using Galaxy Zoo citizen scientists for example. However, this visual classification is not accurate for SRs.) The $2\sigma$ error in F(SR) is less than 0.8\% across all $\log(\Sigma_3^{\rm{rel}})$. In the right hand side panel of \cref{fig:random_sr}, we randomly assign 52\% of all candidate massive SRs to be FRs as well as calculate the measurement errors as before. We recalculate F(SR) for each of the 100 iterations and find that as expected, F(SR) decreases when we account for the misclassification. In the highest density bin, F(SR) decreases by up to a factor of two which is unsurprising considering that massive galaxies are mostly found at high densities. Despite this uncertainty, we can still conclude that there is a trend in F(SR) with $\log(\Sigma_3^{\rm{rel}})$ with a factor $\sim5$ in variation between the two density extremes.

\subsubsection{F(SR) vs $D_{\rm{cen}}$}
There is a close relationship between $\log(\Sigma_3^{\rm{rel}})$ and $D_{\rm{cen}}$ because by definition, the density peaks are close to the centres of groups, and the number density is on average expected to fall off with radial distance from the centre. In \cref{fig:Group_mem_dist}, we plot F(SR) as a function of the projected distance from the group centre, $D_{\rm{cen}}$. We find that for $D_{\rm{cen}} \lesssim 0.12$ Mpc (the first two bins), F(SR) is about 11-14\%. The fraction drops dramatically beyond $D_{\rm{cen}} \sim 0.12$ Mpc after which it settles at about 1\% beyond $D_{\rm{cen}} \sim 0.5$ Mpc. This trend is even stronger than the one seen in \cref{fig:Group_mem_sigma_3_rel} and reveals that massive SRs are strongly biased towards the central $\sim100$ kpc of clusters.

\subsubsection{F(SR) vs $D_{\rm{cen}}^{\rm{norm}}$}
\begin{figure*}
\centering
\includegraphics[width=0.95\textwidth]{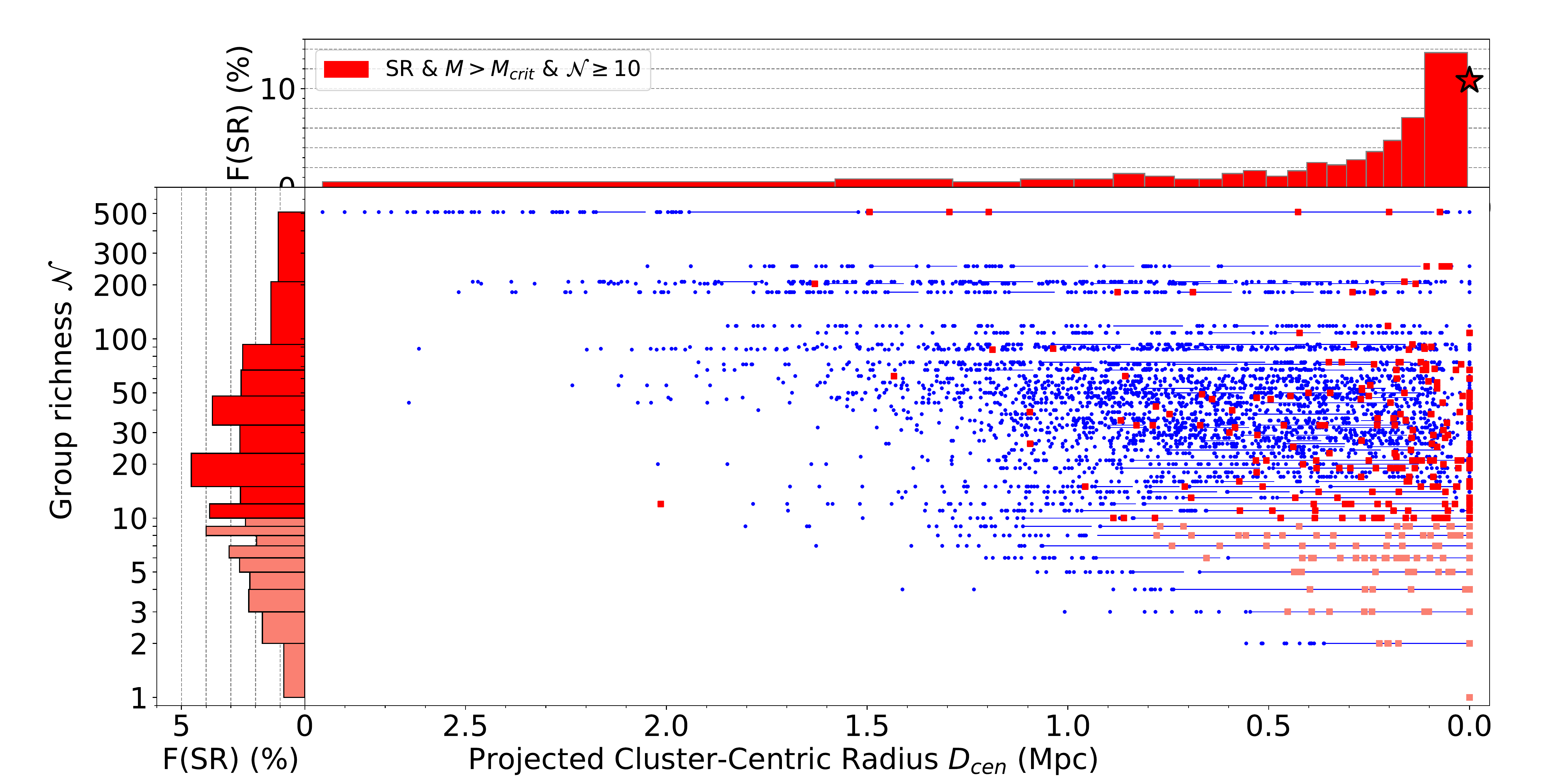}
\caption[Group richness vs projected clustercentric distance]{\textbf{Projected clustercentric distance.} The same as \cref{fig:Group_mem_sigma_3_rel} but now considering the projected comoving clustercentric distance. The red star shows the fraction for galaxies with $D_{\rm{cen}}$ and $\mathcal{N} \geq 10$.}
\label{fig:Group_mem_dist}
\end{figure*}

\begin{figure*}
\centering
\includegraphics[width=0.95\textwidth]{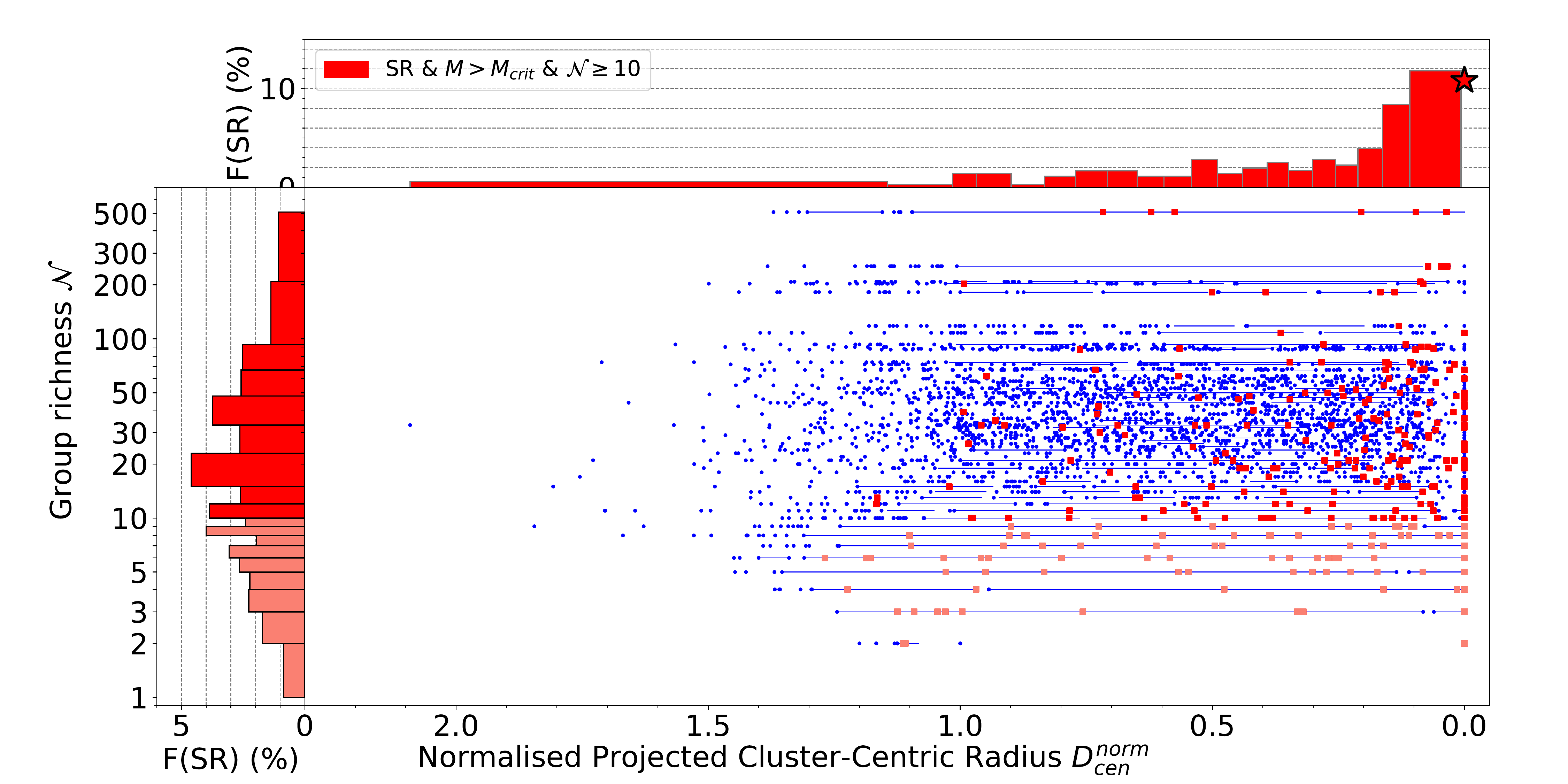}
\caption[Group richness vs normalised projected clustercentric distance]{\textbf{Normalised projected clustercentric distance.} The same as \cref{fig:Group_mem_dist} except that projected distances are normalised to the 90th percentile. The red star shows the fraction for galaxies with $D_{\rm{cen}}^{\rm{norm}}$ and $\mathcal{N} \geq 10$.}
\label{fig:Group_mem_dist_normed}
\end{figure*}

Finally, we normalise the groups to the 90th percentile to bring all groups to a comparable scale in terms of group radius (\cref{fig:Group_mem_dist_normed}). We find that the F(SR)-$D_{\rm{cen}}^{\rm{norm}}$ relation is less skewed but looks similar overall to the $D_{\rm{cen}}$ relation from \cref{fig:Group_mem_dist}. Here we conclude that within the inner 20\% of the group/cluster radius, the fraction of galaxies that are massive SRs is at least 8\%. Outside the same radius, the fraction is less than 4\% for all bins.

\subsection{Disentangling stellar mass and environment}
\label{sec:disentangling}
Here, we examine the relationship between stellar mass and environment using our relative density estimator. In \cref{fig:sigma_rel_fixed_mass}, we plot F(SR) as a function of $\log(\Sigma_3^{\rm{rel}})$ for intervals of fixed stellar mass. Each bin in stellar mass contains about 200 galaxies, and each density bin contains about 40 galaxies. We calculate the error in F(SR) in the same way as in the right panel of \cref{fig:random_sr} i.e. we include the misclassification bias. In all but one stellar mass bin, we see in increase in F(SR) within about $\log(\Sigma_3^{\rm{rel}}) \approx -0.5$ Mpc$^{-2}$. Within the errors, the mass interval $11.5 \leq \log(M) < 11.66$ does not show a strong trend, but F(SR) at $\log(\Sigma_3^{\rm{rel}}) \approx 0$ is about a factor of two greater than at $\log(\Sigma_3^{\rm{rel}}) \approx -1.5$ Mpc$^{-2}$ for the same mass bin. The increase in F(SR) with stellar mass is clearly visible from the difference between the F(SR) profiles for each stellar mass bin.\par

In \cref{fig:mass-radius-dist}, we plot the stellar mass distribution for SRs with $M\geq M_{\rm{crit}}$ (red) as a function of environment (dotted, dashed and solid curves). All other galaxies are contained within the blue curves for the same environments. We find no significant differences between the blue curves suggesting that the stellar mass distribution of FRs and spirals doesn't change as a function of environment. One clear difference is that the peak is more defined for FRs and spirals in clusters compared to the field environment. On the other hand, there is a systematic shift in the SR stellar mass distribution from the field towards the cluster regime. This is consistent with the hierarchical formation scenario where clusters are built up from the mergers of groups, and in parallel the central SRs also build their stellar mass via dry mergers \citepalias{cappellari2016structure}. It is less consistent with the hypothesis that the massive SRs reach the cluster centres by dynamical friction alone (see \autoref{sec:comparisonwork}) as there would not necessarily be a connection between the mass of the BCG and the cluster under that scenario. Note that \cref{fig:mass-radius-dist} does not show that the frequency of SRs is a result of the changing stellar mass distribution, as the red curve is only for massive SRs in the first place. The only free parameter in \cref{fig:mass-radius-dist} is environment.\par
We do not attempt to produce the mass-size relation as a function of environment as we do not have consistent radii across all galaxies. For galaxies in the NSA, but not in MaNGA, we only have Petrosian radii, whereas we measure $R_e$ directly from the photometry (see \citetalias{graham2018angular}). The SDSS Petrosian radii are numerically similar to $R_e$ but are defined in a completely different way. Moreover, there can easily be deblending issues in a dense cluster environment and so the radii are not reliable in the regime where we would require them to be accurate in order to observe the steep trend seen in \cite{cappellari2013effect} and fig. 13 of \citetalias{graham2018angular}.\par
At this point, we have not mentioned anything about possible edge effects where galaxies are close to the boundary of the SDSS footprint (see \citealp{berlind2006groups} where the footprint was much smaller). It is possible that if galaxies are close to the edge, then a portion of its enclosing group could be missing from the catalogue. To assess the frequency of such a scenario, we select a random sample of 50,000 photometric objects that lie with a square of width 10 Mpc centred on each MaNGA galaxy, regardless of their redshift. We also select spectroscopic galaxies within the box that lie within a velocity difference of $\pm$3000 km s$^{-1}$ from each MaNGA galaxy. While the footprint is only well traced by the photometric catalogue which comes from imaging, the spectroscopic galaxies are useful to assess the local density of the MaNGA galaxy. (This analysis is only approximate so we don't use the groups here.) We divide up the squares into 25 smaller squares each with sides 2 Mpc in length. We simply count the number of objects in each squares and count how many squares have 5\% or less of the maximum number of objects in any one square. If 5 or more squares are sufficiently underpopulated, we visually inspect the footprint. We find about 200 MaNGA galaxies ($\sim5\%$) are close to the edge of the footprint. It is non-trivial to estimate what fraction of these galaxies are actually affected, since some will be in genuinely sparse environments (about half of all MaNGA galaxies are isolated). Nevertheless, we redo the figures presented in this section without the groups enclosing these galaxies, and find that our conclusions are unchanged.

%% file: Comparison.tex
\section{Comparison with other work}
\label{sec:comparisonwork}
With the large sample that MaNGA provides, we are able to separate MaNGA galaxies into bins of stellar mass and still retain sufficiently large numbers in each bin to detect meaningful signals which are not affected by random errors in environmental parameters. We now use the MaNGA sample for which we have $\lambda_{R_e}$ to make comparisons with the work of three other studies which have claimed that once stellar mass has been accounted for, there are no residual trends with either the angular momentum proxy $\lambda_{R_e}$ or the fraction of slow rotators as a function of environment. We have already provided clear evidence that $\Sigma_3$ cannot be used to accurately predict the abundance of massive SRs which is highly sensitive to the cluster morphology. Similarly, the fraction of massive SRs is not a simple function of group richness. While we do not necessarily have access to the same datasets or parameters, we can nevertheless still make a qualitative comparison in order to ``test our machinery''.\par
We choose not to use the volume-weighted sample, which is equivalent to a volume-limited survey, because massive SRs are vastly outnumbered by FRs in such a sample. Moreover, large weights are given to very unusual objects and so our statistics will be dominated by these objects. The weights are defined in terms of redshift, as galaxies which are observed at low redshifts are upweighted and vice versa. However, as the MaNGA selection function requires that galaxies of increasing luminosity (and therefore stellar mass) are observed at higher redshifts, the weights also depend on stellar mass. Hence, the range of possible weights is restricted for bins of fixed mass, and so for the regime where genuine SRs exist, the effect of using the volume weighted sample is in any case minimal.

\subsection{Comparison with Veale et al. 2017}
\label{sec:veale}
In \cref{fig:veale}, we aim to reproduce fig. 8 from \cite{veale2017angular} which was used to argue for the absence of strong trends in angular momentum with three different environmental parameters. We keep the same mass bins, simply for direct comparison, despite the fact that the middle mass bin ($10.9<\textrm{log}(M_*)<11.7$) straddles the critical mass above which SRs are dominant. This implies that the \cite{veale2017angular} sample includes some galaxies that are likely ``spurious'' SR, in the sense of not being dry merger relics. Furthermore, the number counts are not equally shared between these three stellar mass bins, which means that the uncertainty is not the same for each bin.\par

\begin{figure*}
\centering
\includegraphics[width=0.95\textwidth]{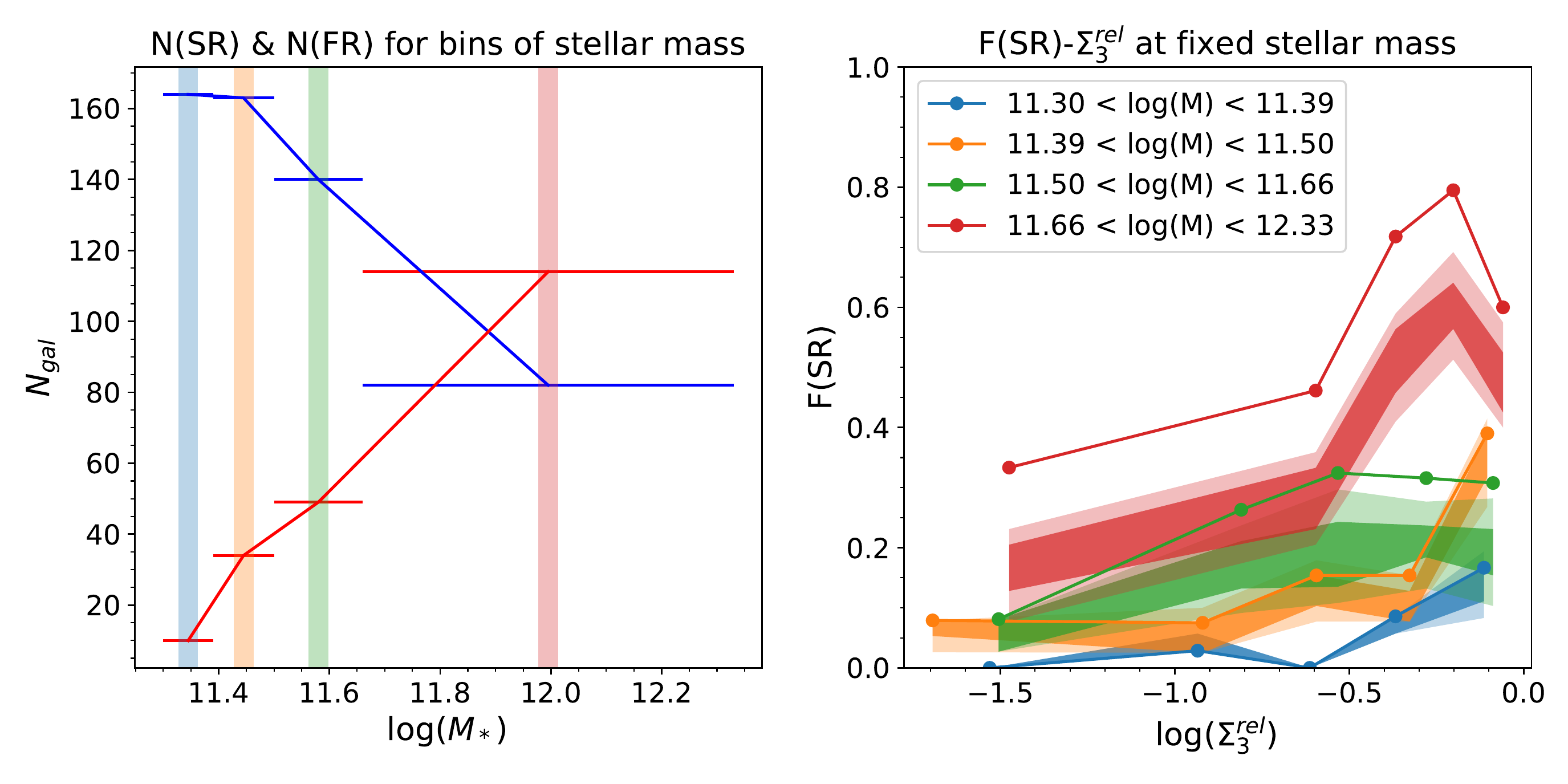}
\caption[F(SR) vs $\Sigma_3^{\rm{rel}}$ at fixed stellar mass.]{\textbf{F(SR) vs $\Sigma_3^{\rm{rel}}$ at fixed stellar mass.} \textit{Left}: Stellar mass distribution of SRs and FRs above $M_{\rm{crit}}$. Each bin has equal numbers of galaxies. \textit{Right}: For each bin shown at top, we plot the F(SR) as a function of $\log(\Sigma_3^{\rm{rel}})$. We also show each relation including the errors in $\lambda_{R_e}$ and $\epsilon$ as well as accounting for the misclassification bias.}
\label{fig:sigma_rel_fixed_mass}
\end{figure*}

We do not share any environmental parameters with \cite{veale2017angular}. Instead of the halo mass, we use the group richness $\mathcal{N}$ as a proxy. In place of the luminosity density within the 10th nearest neighbour, we use the projected surface density within the 10th nearest neighbour, $\log(\Sigma_{10})$. We compare values of $\log(\Sigma_{10})$ (units of $\textrm{ Mpc}^{-2}$) and $\log(\nu_{10})$ (units of $L_{{\odot}K} \textrm{ Mpc}^{-3}$) from table 2 of \cite{cappellari2011atlas3db} and find that they vary almost linearly with reasonable amounts of scatter. We do not have an equivalent for the overdensity $\delta_g$, which measures environmental density on large scales of a few Mpc and is related to the halo mass \citep{veale2017angular}. Instead, we swap $\delta_g$ for $\log(\Sigma_3)$ which is our local density estimator. However, this means that the middle columns of our plots are not directly comparable, as they probe completely different scales. For the three mass bins (referred here as low-mass, intermediate-mass and high-mass bins), we plot $\lambda_{R_e}$ as a function of the three environmental parameters (upper panels of \cref{fig:veale}). Although \cite{veale2017angular} calculated the mean for three environment bins (referred here as low-density, intermediate-density and high-density bins), we choose to calculate the median which is less affected by outliers. We also plot F(SR) for each bin in density and stellar mass (lower panels of \cref{fig:veale}).\par
We fail to detect any strong trends in $\lambda_{R_e}$ as a function of either of the three environmental parameters for the lowest mass bin. However, for the intermediate mass bin, there is a visible trend in that $\lambda_{R_e}$ decreases with environmental density. Perhaps the strongest trend is seen in the median for the intermediate-mass distribution as a function of $\log(\Sigma_3)$. There is a similar trend in the highest mass bin but it is slightly weaker. However, the dominance of stellar mass is clearly visible, as the difference \textit{between} the median lines is much greater than the gradient of the lines themselves. Each distribution is highly asymmetric and the inner $1\sigma$/$2\sigma$ percentiles would be far from the median lines if we were to show them. (For the sake of clarity, we omit these lines.) As the slope of the gradients are upper limits, they would be even weaker for a volume weighted sample.\par
There is a marked difference in the histograms of F(SR) as a function of environment  between the stellar mass bins (but not between the three environmental parameters which show similar distributions). First we consider the bottom left panels of our \cref{fig:veale} and fig. 8 of \cite{veale2017angular}. Our results show some general similarity. \cite{veale2017angular} find an increase of $\sim20\%$ for F(SR) between the low-mass halos and high-mass halos. In our data, we see a $\sim30\%$ for F(SR) for comparable environments. For the intermediate- and low-mass bins,  \cite{veale2017angular} find a constant F(SR) as a function of halo mass. We agree for the lower-mass bin, and we find a slight increase F(SR) as a function of halo mass for the intermediate mass bin.\par
Next, we consider the bottom right panels of both figures. The trends for the high mass bin are not unlike those seen for $M_{\rm{halo}}$, in that we see a $\sim30\%$ increase in F(SR) between bins of low and high density, where \cite{veale2017angular} see a $\sim20\%$ for a similar regime. For the intermediate mass bin, we actually find a shallower trend compared to \cite{veale2017angular}. Meanwhile, we agree on the slightly decreasing trend for the lowest mass bin. However, as previously mentioned, these are not the relics of dry mergers but are low-mass galaxies that are slowly rotating.\par

\begin{figure}
\centering
\includegraphics[width=0.45\textwidth]{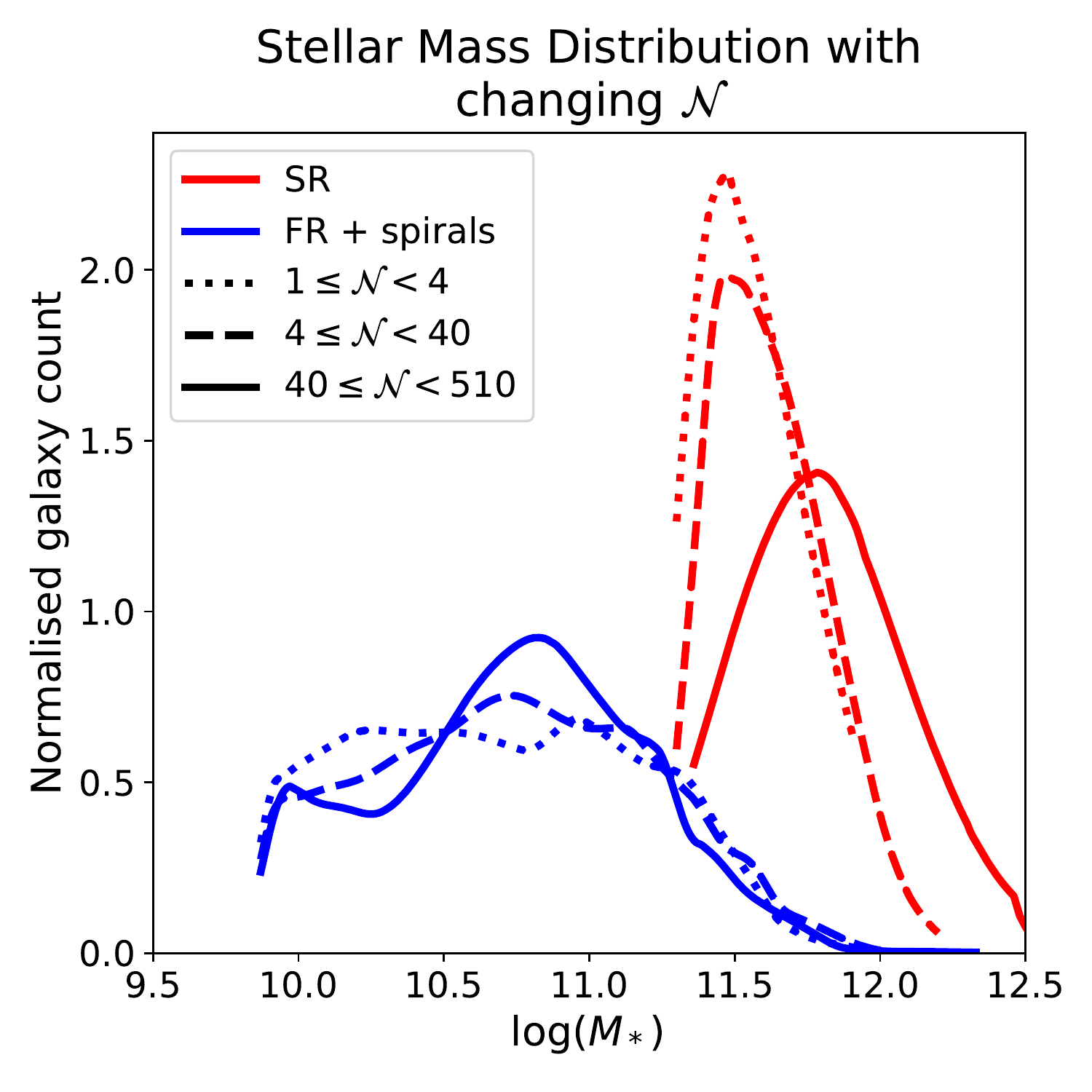}
\caption[Stellar mass distributions as a function of environment.]{\textbf{Stellar mass distributions as a function of environment.} We plot probability distributions for massive SRs in red and all other galaxies in blue. We split each population into environment bins representing the field (dotted), group environments (dashed) and clusters (solid). We truncate the blue curves below the completeness limit of $\log(M)=9.87$.}
\label{fig:mass-radius-dist}
\end{figure}

Finally, we consider the middle bottom panel of \cref{fig:veale}. We see a strong increase in F(SR) of $\sim40\%$ between the low and high density ($\Sigma_3$) bin. The trends for the intermediate- and low-mass bins are similar to the $\Sigma_{10}$ F(SR) histograms. The lack of high-mass galaxies at $\log(\Sigma_3) \gtrsim 2$ in MaNGA can be explained as a result of the fact that there is an intrinsic maximum density that massive galaxies which are large in radius can find themselves.\par
Overall, we find stronger increasing trends for F(SR) as a function of environment for the highest mass bin compared to those found by \cite{veale2017angular}. We conclude from \cref{fig:veale} at for massive galaxies, environment \textit{does} influence the abundance of dry merger relics (see \cref{sec:discussion}).\par
The histograms are relatively unaffected by the volume weighting because in fixed stellar mass bins, the weights will not vary greatly. As a result, the net effect of switching to the volume-weighted sample is that F(SR) decreases largely uniformly across all parameters by up to 20\%. The trends themselves are preserved and so we would draw the same conclusions for the volume-weighted MaNGA sample.\par
 
\begin{figure*}
\centering
\includegraphics[width=0.9\textwidth]{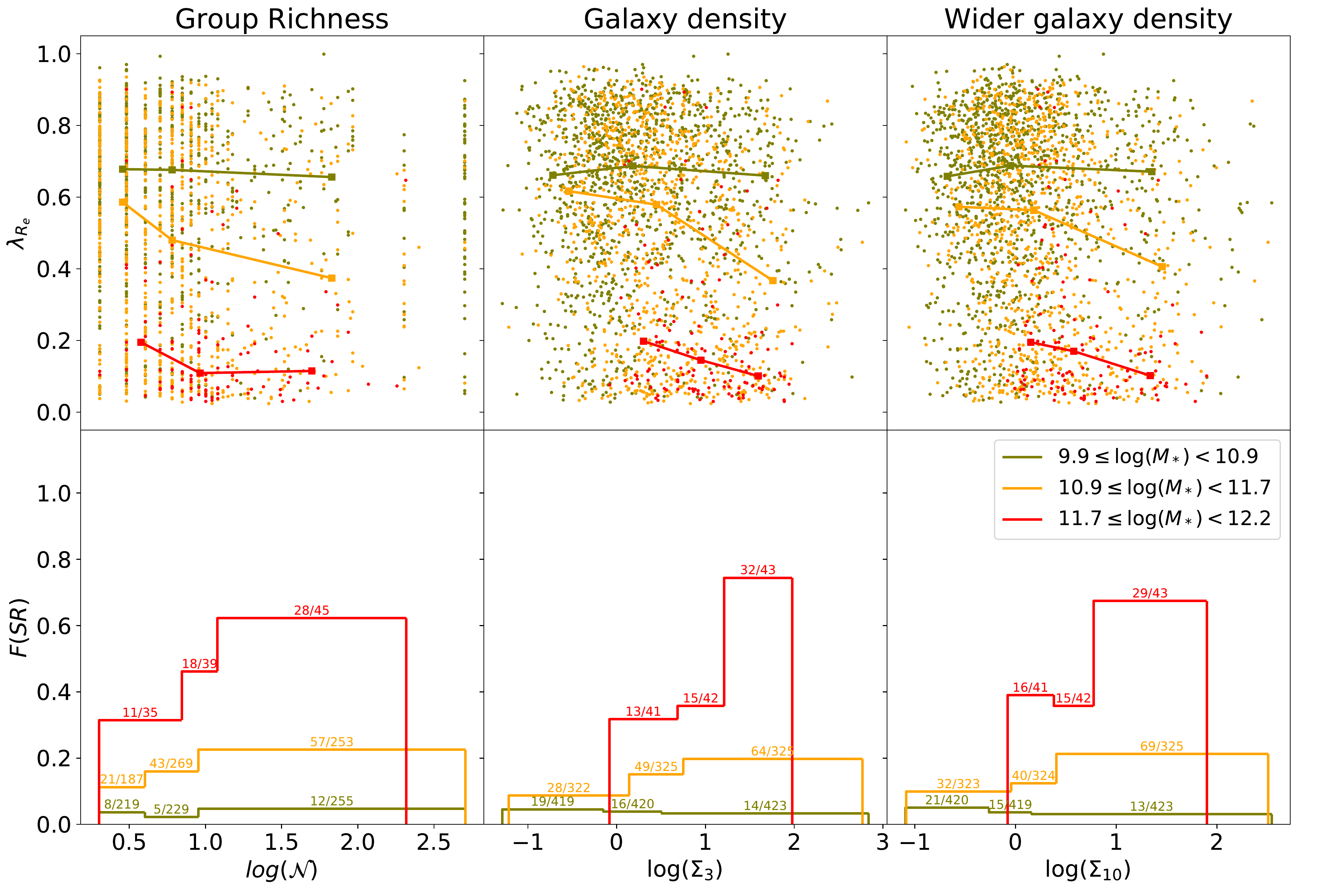}
\caption[Comparison with Veale et al. 2017]{\textbf{Comparison with \cite{veale2017angular}.} \textit{Top}: We plot $\lambda_{R_e}$ as a function of $\log(\mathcal{N})$ (left), $\log(\Sigma_3)$ (middle) and $\log(\Sigma_{10})$ (right). As in fig. 8 of \cite{veale2017angular}, we split the sample into bins of stellar mass: $9.9 \leq \log(M) < 10.9$ (green), $10.9 \leq \log(M) < 11.7$ (olive) and $11.7 \leq \log(M) < 12.2$ (red). For each environmental parameter, we plot the median $\lambda_{R_e}$ for bins of equal number. \textit{Bottom}: The same as above except for the fraction of SRs, $f_{\rm{slow}}$. We give the number of SRs and total galaxies in each bin.}
\label{fig:veale}
\end{figure*}

\subsection{Comparison with Greene et al. 2017}
\begin{figure*}
\centering
\includegraphics[width=0.9\textwidth]{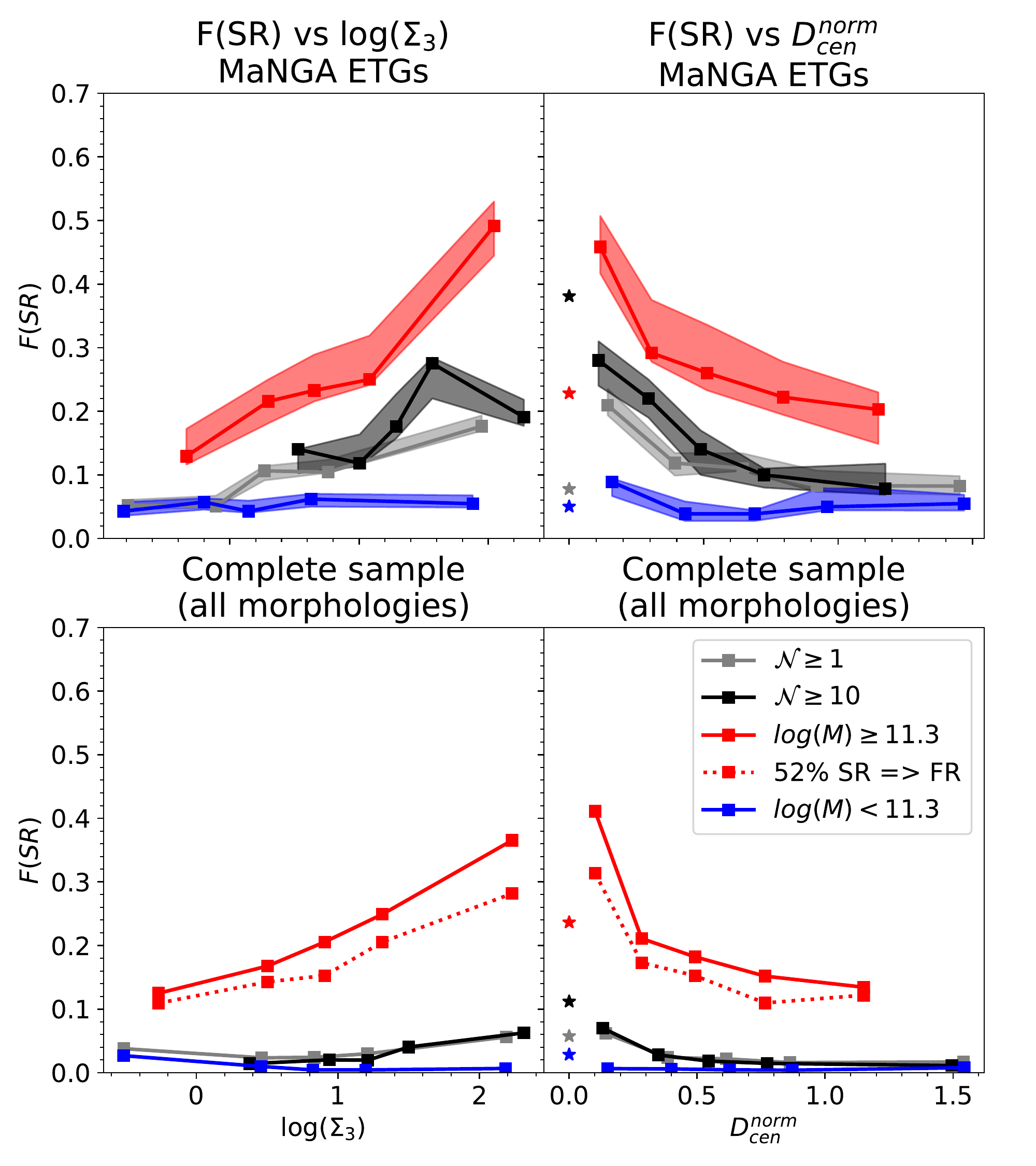}
\caption[Comparison with Greene et al. 2017]{\textbf{Comparison with \cite{greene2017kinematic}.} \textit{Top Left}: F(SR) as a function of $\log(\Sigma_3)$. We show trends for all galaxies (grey) galaxies in groups of 10 or more members (black) and galaxies more and less massive than $M_{\rm{crit}}$ (red and blue respectively) for all group memberships. Each subsample is split up into five bins of equal numbers of galaxies. \textit{Top Right}: F(SR) as a function of $D_{\rm{cen}}^{\rm{norm}}$ normalised to the 90th percentile for MaNGA ETGs only. The stars correspond to the values at $D_{\rm{cen}}^{\rm{norm}} = 0$ for each subsample. \textit{Bottom Left}: As top left except for the complete volume-limited sample (all morphologies). As this sample contains SR candidates, we randomly assign FR classifications to 52\% of our SR candidates (dashed-red). \textit{Bottom Right}: As top right except for the complete sample (all morphologies).}
\label{fig:greene}
\end{figure*}

\cite{greene2017kinematic} studied the fraction of SRs in a sample of 538 ETGs ($M \geq 10^{10} \textrm{ M}_{\odot}$) taken from the MPL-5 sample as a function of normalised cluster radius. They used the group catalogue of \cite{yang2007groups} based on SDSS DR7 that assigns galaxies to groups using a friends-of-friends algorithm before assigning a halo mass based on the total galaxy luminosity and refining using an iterative procedure. Hence, they had estimates for the virial radii of each cluster which we do not have. To make a comparable study, we take the radius of each group to be the 90\% percentile, as opposed to the maximum which may be affected by outliers. \cite{greene2017kinematic} used $\log(\Sigma_5)$ which is not identical but is tightly correlated with $\log(\Sigma_3)$ which we use here. In the top panels of \cref{fig:greene}, we show the fraction of SRs as a function of $D_{\rm{cen}}^{\rm{norm}}$ (left) and $\log(\Sigma_3)$ (right) for four subsamples: All ETGs, all ETGs in groups with $\mathcal{N} \geq 10$, ETGs less massive than $M_{\rm{crit}}$, and ETGs more massive than $M_{\rm{crit}}$. We note that we use a different mass cut, and specifically we use $M_{\rm{crit}}$, which defines the minimum mass of genuine dry-merger SRs and is more physically motivated than the two used by \cite{greene2017kinematic}.\par
We find that F(SR) for the low mass sample is less than 10\% for all values of $D_{\rm{cen}}^{\rm{norm}}$, but we stress that these are essentially mis-classified SRs (such as face-on counter-rotating disks), so their trend is not very meaningful. We show it simply to be consistent with \cite{greene2017kinematic}. For the high mass sample, the fraction increases from about 20\% at the group boundary to $\sim45\%$ at the group centre. This is in clear contrast with that shown in fig. 2 of \cite{greene2017kinematic} where there is essentially no trend with normalised group radius. Although we show the relation for all galaxies in all groups, $D_{\rm{cen}}^{\rm{norm}}$ is essentially meaningless for small groups ($D_{\rm{cen}}^{\rm{norm}} = 0$ for isolated galaxies). Hence, we also show the relation considering only groups with 10 or more members as in \cref{fig:Group_mem_dist_normed} etc. Unsurprisingly, this relation lies between the low-mass and high-mass relation, and show an increasing trend with decreasing $D_{\rm{cen}}^{\rm{norm}}$. All the trends with $D_{\rm{cen}}^{\rm{norm}}$ become flatter with volume weighting (apart from the low-mass trend which is already flat without volume weighting), with only the high-mass bin retaining any noticeable trend.\par
There is a strong linear trend visible for the high mass sample where F(SR) increases from $\sim15\%$ to $\sim50\%$ for massive MaNGA SRs between $\log(\Sigma_3) \approx -0.2$ and $\log(\Sigma_3) \approx 2.2 \textrm{ Mpc}^{-2}$. This trend contrasts with that seen in fig. 1 of \cite{greene2017kinematic} where F(SR) is constant as a function of $\Sigma_5$ within the errors. $\log(\Sigma_3)$ is restricted for $\mathcal{N} \geq 10$ (see \cref{fig:Group_mem_sigma_3}), and here we do not see any strong trend for all galaxies for these groups. However, this is because high-mass galaxies are greatly outnumbered by low-mass galaxies.\par
To check that we are not affected by the MaNGA selection function, we plot F(SR) for the volume-limited sample using the same selection criteria in the bottom two panels of (\cref{fig:greene}). We find qualitatively the same result with a slight reduction in F(SR) for the high mass sample.\par

\subsection{Comparison with Brough et al. 2017}
Finally, we carry out a comparable analysis to that shown in Fig. 11 of \cite{brough2017kinematic} which looks at the mean deprojected $\lambda_{R_e}$ for fixed density as a function of stellar mass, and at fixed stellar mass as a function of density. The deprojection is achieved by assuming that the demarcation line between FRs and SRs used in \cite{emsellem2011atlas3d}, $\lambda_{R_e} = 0.31\sqrt{\epsilon}$, approximately separates ETGs by their \textit{intrinsic} edge-on angular momentum. $\lambda_{R_e}$ certainly should not be interpreted without knowledge of $\epsilon$ (as we have done in \cref{sec:veale}), but as more data have become available, the demarcation line has been refined (see \cref{fig:lambda_ellip} and \citetalias{cappellari2016structure}). Furthermore, \cite{brough2017kinematic} deproject SRs which are not intrinsically axisymmetric and so the deprojection is not meaningful for these galaxies. (Even if we observed all SRs ``edge-on'' i.e. $\epsilon \sim 0.3-0.4$, we would still observe low $\lambda_{R_e}$.) Finally, it is clear that deprojecting galaxies of low $\epsilon$ and high $\lambda_{R_e}$ using $1/\sqrt{\epsilon}$ will result in unphysical values (see \citealp{greene2017kinematic} and fig. 6 of \citealp{lee2018environment}).\par
In \cref{fig:lambda_ellip}, we plot how the edge-on ($i=90\degree$ ) theoretical (magenta) line changes as a function of inclination (\citealp{binney2005rotation}, \citealp{cappellari2007sauron}, \citetalias{cappellari2016structure}). We also plot lines which show how a galaxy with a given $(\lambda_{R_e}^{\rm{intr}},\epsilon^{\rm{intr}})$ at $i=90\degree$ moves across the diagram as the inclination progresses towards face on ($i=0\degree$; \cite{cappellari2007sauron}). We could use these lines to deproject $\lambda_{R_e}$ and $\epsilon$ to the intrinsic values for FRs. However, the inclination is generally undefined due to measurement errors in $\epsilon$, and measurement errors of $\sim0.05$ in $\lambda_{R_e}$ can easily change the deprojected $\lambda_{R_e}^{\rm{intr}}$ by a larger amount, especially at low $\epsilon$. Finally, the functional forms that predict the magenta line and its projections are at a far higher precision than is generally required, especially considering the measurement errors.\par
In light of this, we derive a simple empirical function that can predict the deprojected value of $\lambda_{R_e}$ to within an accuracy of about 0.05. For all SRs or galaxies with $\epsilon \geq 0.4$, we assume that the observed $\lambda_{R_e}$ is the edge-on one, and we do not correct for any inclination effects. For FRs rounder than $\epsilon = 0.4$, we calculate the edge-on, intrinsic $\lambda_{R_e}^{\rm{intr}}$ using an empirical correction for inclination
\begin{equation}
\lambda_{R_e}^{\rm{intr}}-\lambda_{R_e}=0.16\textrm{exp} \Bigg [{-\frac{(\lambda_{R_e} - 0.4)^2}{0.04}}\Bigg]x + 0.2 x^2\textrm{,}
\label{eq:inc_corr}
\end{equation}
where
\begin{equation}
x =  \frac{0.4-\epsilon}{0.4}\textrm{,}
\label{eq:inc_corr2}
\end{equation}
and $\epsilon$ and $\lambda_{R_e}$ are the apparent ellipticity and beam corrected stellar angular momentum parameter respectively.\par
\cref{eq:inc_corr} has two components on the right hand side. The first component models the change in $\lambda_{R_e}$ as a function of $\epsilon$ and $\lambda_{R_e}$. The form is a Gaussian which is scaled according to $\epsilon$, and peaks at $\lambda_{R_e}=0.4$, which is roughly where the inclination lines are at their maximum gradient. The change is greatest at intermediate $\lambda_{R_e}$ and falls off towards the two extremes of $\lambda_{R_e}=0$ and $\lambda_{R_e}=1$. When $\epsilon=0$ and $\lambda_{R_e}=0.4$, the change in $\lambda_{R_e}$ due to this component is simply 0.16. The second component, proportional to $x^2$, models the change in $\lambda_{R_e}$ as a function of $\epsilon$, assuming that the increase in $\lambda_{R_e}$ is proportional to the square of the fractional distance $x$ from $\epsilon=0.4$ in the range $0 \leq \epsilon \leq 0.4$. In other words, when $\epsilon=0$ i.e. maximum distance from $\epsilon=0.4$, the component evaluates to 0.08, and when $\epsilon=0.4$, the component evaluates to zero.\par
\begin{figure}
\centering
\includegraphics[width=0.49\textwidth]{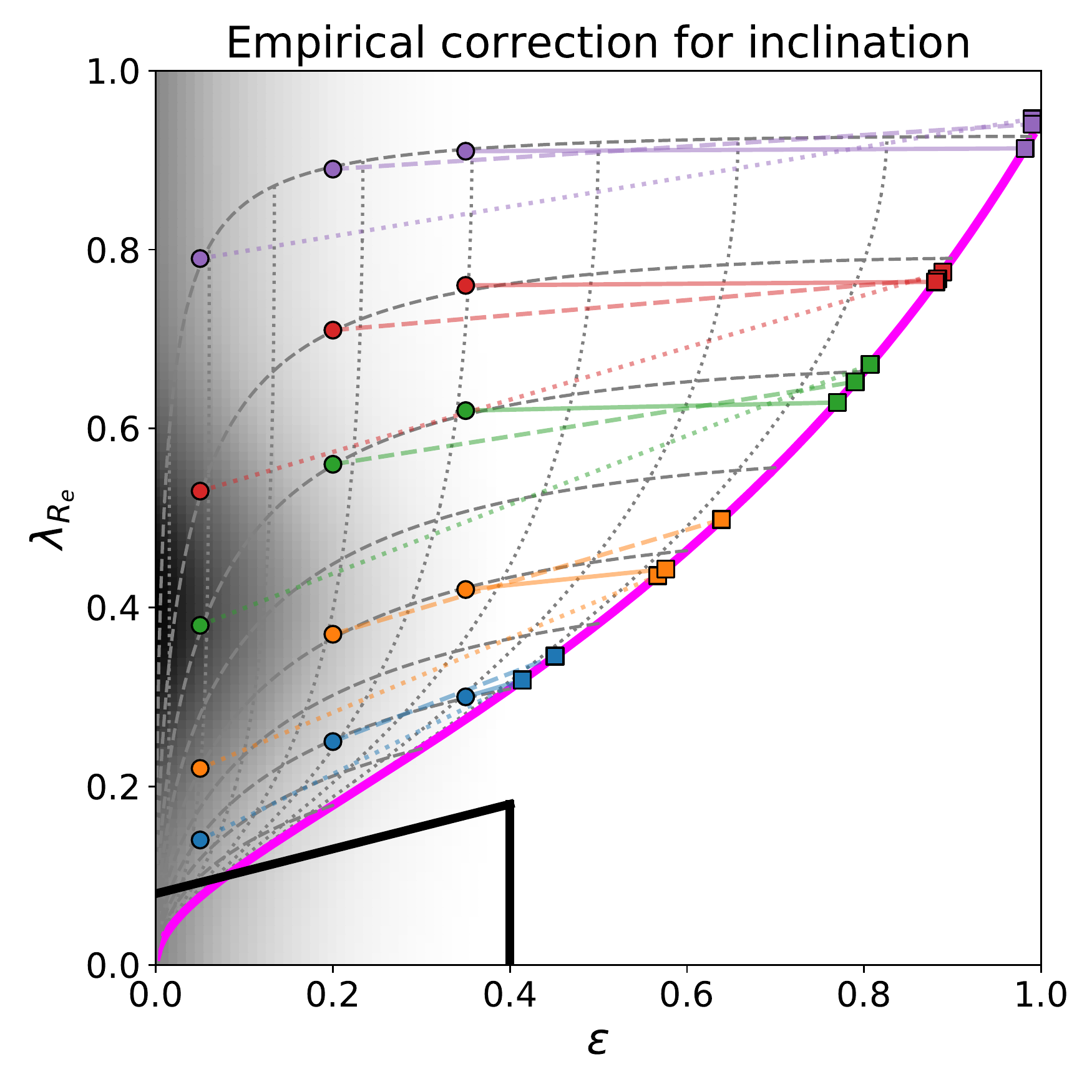}
\caption[Correcting for inclination]{\textbf{Correcting for inclination}. Illustration of our empirical inclination correction (\cref{eq:inc_corr}). The magenta, black and grey dashed lines are the same as in \cref{fig:lambda_ellip}. The coloured circles are the apparent positions for 15 mock galaxies. The colours correspond to a particular intrinsic $\epsilon^{\rm{intr}}$ to guide the eye. For each colour, three values of $\epsilon$ are chosen as 0.05, 0.2 and 0.35, and $\lambda_{R_e}$ is chosen so that the point lies on one of the $\epsilon^{\rm{intr}}$ inclination tracks. The inclination corrected locations are shown as coloured squares, and each pair of points is joined by a line of the same colour. We do not calculate $\epsilon^{\rm{intr}}$ but simply match up the corresponding value on the magenta line. The black shaded region indicates the magnitude of the inclination correction at each position on the plane. The peak is centred at $(\lambda_{R_e},\epsilon)=(0.4, 0)$, and there is no correction for galaxies flatter than E4.}
\label{fig:inc_corr}
\end{figure}

We illustrate \cref{eq:inc_corr} in \cref{fig:inc_corr}, where we plot the apparent $(\lambda_{R_e}, \epsilon)$ and deprojected $(\lambda_{R_e}^{\rm{intr}}, \epsilon^{\rm{intr}})$ for 15 mock galaxies. We do not calculate $\epsilon^{\rm{intr}}$ but simply match it with the value of $\lambda_{R_e}^{\rm{intr}}$. (We do this purely for illustration as we are not interested in obtaining $\epsilon^{\rm{intr}}$.) We can assess the performance of the inclination correction by comparing the distance along the magenta line between each coloured square to its corresponding inclination track (grey dashed lines). We find that the correction can predict $\lambda_{R_e}^{\rm{intr}}$ to within about 0.05 which is sufficiently accurate for our purposes. The function was calibrated using the 15 mock galaxies shown, and so performs as intended when $\epsilon \geq 0.05$. The accuracy will decrease below $\epsilon \geq 0.05$ where small shifts in $\lambda_{R_e}$ or $\epsilon$ can would result in large shifts in the deprojected values, but we expect the accuracy to still be within about 0.1 for this $\epsilon$ range.

\begin{figure*}
\centering
\includegraphics[width=0.95\textwidth]{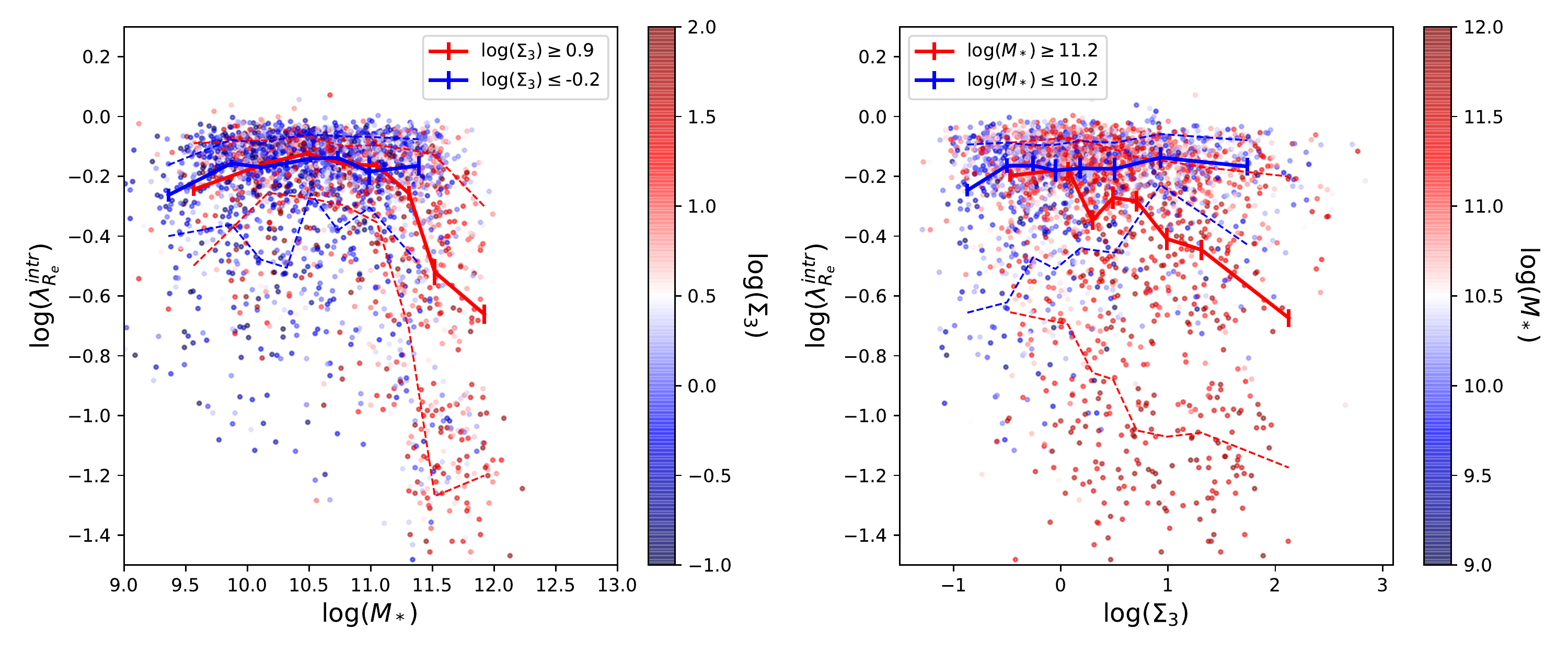}
\caption[Comparison with Brough et al. 2017]{\textbf{Comparison with \cite{brough2017kinematic}}. \textit{Top}: Log of the inclination corrected $\lambda_{R_e}^{\rm{intr}}$ corrected using \cref{eq:inc_corr} as a function of stellar mass for the MaNGA sample. The colours of the points correspond to $\log(\Sigma_3)$ as shown in the colourbar. The red and blue solid lines are the median $\log(\lambda_{R_e}^{\rm{intr}})$ for the upper and lower 25th density percentiles. The lower and upper limits of the upper and lower percentiles respectively are indicated in the colour bar by two solid black lines. For each density percentile, we indicate the extent of the inner $1\sigma$ percentile of the distribution with dashed lines. \textit{Bottom}: Same as above except that the independent variable is the surface density and the colours indicate the stellar mass as shown in the colourbar.}
\label{fig:brough}
\end{figure*}

In the upper panel of \cref{fig:brough}, we plot $\log(\lambda_{R_e}^{\rm{intr}})$ calculated using \cref{eq:inc_corr} as a function of stellar mass for the MaNGA sample. We also plot the median and inner $1\sigma$ percentiles of $\log(\lambda_{R_e}^{\rm{intr}})$ for the lower and upper $\log(\Sigma_3)$ quartiles, shown as blue and red lines respectively. For the low density bin, there is essentially no trend in the intrinsic $\lambda_{R_e}$ as a function of stellar mass, which is to be expected if there are by definition no environmental effects for galaxies in regions of low density. We do not observe the dip in $\log(\lambda_{R_e}^{\rm{intr}})$ seen in the low density-high stellar mass bin in \cite{veale2017angular}. This is simply because very few galaxies of masses greater than the critical mass for SRs reside in low density environments. We do confirm the sharp downward trend for the high density bin at $M\approx M_{\rm{crit}}$, which is the mass above which genuine dry-mergers slow rotators start appearing (\citetalias{cappellari2016structure}).\par
In the lower panel of the same figure, we plot $\log(\lambda_{R_e}^{\rm{intr}})$ as a function of $\log(\Sigma_3)$ for the lower and upper quartiles in stellar mass. We find no trend in the lower mass quartile as expected. However, we see a decrease in the median $\lambda_{R_e}^{\rm{intr}}$ for the high mass bin as $\log(\Sigma_3)$ increases. The distribution in $\log(\lambda_{R_e}^{\rm{intr}})$ for this stellar mass quartile is highly asymmetric, however it is clear that the frequency of galaxies with $\lambda_{R_e} \approx 0.1$ increases with increasing $\Sigma_3$ as evidenced by the distribution of the points themselves. We have a larger sample than \cite{brough2017kinematic} and a robust estimate of $\Sigma_3$ so we are able to reveal trends that have previously been missed. Furthermore, we have a more accurate inclination correction that does not overestimate $\lambda_{R_e}^{\rm{intr}}$ at low $\epsilon$. In retrospect, it would have been better to produce this plot using $\Sigma_3^{\rm{rel}}$ instead of $\Sigma_3$, as the red median line in the lower panel of \cref{fig:brough} would be stronger (compare \cref{fig:Group_mem_sigma_3} with \cref{fig:Group_mem_sigma_3_rel}). However, we would have to reduce the sample size to groups of 10 members or larger so we would have to compromise on accuracy.

%% file: Discussion.tex
\section{Discussion}
\label{sec:discussion}
The results presented in this paper allow us for the first time to provide a fully consistent picture of different results on the importance of environment in influencing the internal kinematics of galaxies. 

\begin{enumerate}
\item Our main conclusion is that the prevalence of massive SRs is closely related to the density \textit{relative} to the peak density of their enclosing clusters, and is not predicted accurately by the absolute galaxy surface density $\Sigma_3$.
\item As a result of the previous point, combining multiple clusters in a single kT-$\Sigma$ relation effectively washes out the individual F(SR) profiles that are clearly observed within clusters.
\item At fixed stellar mass, there \textit{is} a dependence of F(SR) on the environment as traced by $\Sigma_3^{\rm{rel}}$.
\item The mean of the stellar mass distribution for core SRs increases from the field to group environments and on to cluster environments. The same cannot be said for the mass distribution of FR ETGs and spirals.
\end{enumerate}

Our results are fully consistent with the evolutionary picture outlined in sec. 6 of \citetalias{cappellari2016structure}. The BCGs of Coma and the other clusters in our sample (see \citetalias{graham2019dclusters}) are strongly biased towards the densest regions of each particular cluster which encloses them. The fact that the mass of the BCG depends on the cluster richness suggests that the evolutionary history of the most massive SRs is closely linked to the formation of their enclosing clusters. When two clusters merge, the massive SRs also merge on radial orbits and increase their mass and size dramatically. The FRs (and spirals) are unlikely to merge due to their relative velocities.\par
We observe a peak in the distribution of SRs as a function of group richness. This presence of such a peak at intermediate richness is in agreement with the accepted view that galaxy groups, rather than more massive clusters, are the most likely environments for galaxy interactions to occur because the velocity dispersion of the group is near that of the velocities of the galaxies themselves \citep{barnes2006groups, zabludoff1998groups, hashimoto2000environment, wilman2005groups, martinez2006groups}. Groups of 20 or 30 members are the progenitors of the massive clusters, and as the SRs merge, their frequency decreases.\par
We are able to reproduce the kT-$\Sigma$ relation for SRs where in general, the fraction of galaxies in bins of density that are massive SRs increases with increasing local number density (\cref{fig:Group_mem_sigma_3}). Our relation is the first to only count SRs if their stellar mass is greater than $\log(M)=11.3$, which has been shown to effectively separate genuine core SRs from galaxies that exhibit low angular momentum for other reasons, such as slowly-rotating low-mass irregular galaxies (\citealp{cappellari2013effect}; see fig. 13 of \citetalias{graham2018angular}). The genuine SRs are believed to be the true dry merger relics and would show a core surface brightness profile (\citealp{krajnovic2013atlas3d}, \citetalias{cappellari2016structure}, \citealp{krajnovic2019cores}) if we had HST imaging for those galaxies. Thus, it is crucial that we do not include low mass SRs when calculating F(SR), as they cannot have formed by dry mergers.\par
We find that F(SR) is only a weak function of $\Sigma_3$, increasing by at most only 5\% across a decade of increasing number density. In contrast, F(SR) jumps by about 15\% between the ATLAS$^{\rm{3D}}$ field sample and the core of Virgo, although \cite{cappellari2011atlas3db} did not specify a minimum critical mass to define SRs. The difference between F(SR) as a function of $\Sigma_3$ (\cref{fig:Group_mem_sigma_3}) and F(SR) as a function of $\Sigma_3^{\rm{rel}}$ (\cref{fig:Group_mem_sigma_3_rel}) is striking. In general, the massive SRs show a clear preference towards the densest regions of their respective clusters, regardless of the group richness. We have confirmed the hypothesis outlined in \citetalias{graham2019bcatalogue} regarding the fact that F(SR) profiles for individual clusters can be smoothed out and flattened when combined with other clusters. In fact, this effect is seen to be strong in our case as we have hundreds of groups where the maximum local number density between groups can vary considerably.\par
The peak in the F(SR)-$\log(\Sigma_3^{\rm{rel}})$ relation appears to not be at $\log(\Sigma_3^{\rm{rel}}) = 0$ precisely, but is offset at about $\log(\Sigma_3^{\rm{rel}}) = -0.2$. This is because the peak density is sometimes not at the most massive ETG, or even at the group centre. In fact, very close pairs between low-mass galaxies in dense environments can mean that the galaxy at the precise peak density is largely unremarkable. As $D_3$ approaches zero, $\log(\Sigma_3^{\rm{rel}}) \equiv \log[3/(\pi D_3^2)]$ increases with increasing rapidity. This is the case for the most massive clusters in our sample where the massive SRs are not at the peak density. In fact, the distance between two galaxies in a close pair can be smaller than the \textit{angular size} of the most massive ETGs. However, we find that this only affects the most massive ETGs which are also the largest (see \autoref{fig:sigma_rel_fixed_mass}), as more lightweight SRs are smaller in size. Hence, the slight offset of the peak seen in $\log(\Sigma_3^{\rm{rel}})$ is likely an observational artifact. \textit{Furthermore, as the clusters are projected, we would likely find that the massive SRs are genuinely sitting at the peak density if we were to view the cluster from all possible viewing angles.}\par
As expected, \cref{fig:Group_mem_dist} looks qualitatively similar to \cref{fig:Group_mem_sigma_3_rel}, given that in some sense, $D_{\rm{cen}}$ is also a relative measurement (comparing each galaxy with the central galaxy where $D_{\rm{cen}} = 0$). In fact, F(SR) as a function of $D_{\rm{cen}}$ is even more skewed towards the minimum value than $\Sigma_3^{\rm{rel}}$, suggesting that the projected distance is a better predictor of the abundance of massive SRs than even $\Sigma_3^{\rm{rel}}$. Indeed, the median distance for the massive SRs is less than 0.2 Mpc compared with nearly 0.6 Mpc for the FRs. Unlike friends-of-friends methods which choose the most luminous galaxy to be the centre of the group, our centre is taken as the peak of the underlying galaxy distribution and so is independent of the stellar mass. The fact that the peak is close to zero indicates the robustness of the F(SR)-$D_{\rm{cen}}$ relation shown here.\par
In \cref{sec:disentangling}, we investigate the hypothesis that the kT-$\Sigma$ relation is a result of the changing stellar mass distribution with environment by calculating F(SR) for bins of $\log(\Sigma_3^{\rm{rel}})$ at fixed stellar mass intervals (\cref{fig:sigma_rel_fixed_mass}). We find that F(SR) generally increases with relative density although each trend is not particularly strong. However, the presence of such a trend again rules out the scenario where SRs are at the cluster cores simply as a result of dynamical friction. If dynamical friction were indeed the cause, then we would expect the F(SR) profiles to be flat, as dynamical friction is only dependent on the stellar mass \citep{scott2014distribution}.\par
We also extend the analysis of \cite{cappellari2013effect} who compared the mass-size distributions of SRs and FRs in the field and in the Coma cluster. In \cref{fig:mass-radius-dist}, we compare the stellar mass distributions in the field, group and cluster environments for SRs above $M_{\rm{crit}}$ and all other galaxies. The systematic increase in the SR stellar mass distribution from the field to the cluster supports the hypothesis that clusters are built up in a hierarchical formation process from smaller groups, as this growth is mirrored in the merging of SRs initially at rest at the centre of the subcluster potentials \citep{cappellari2016structure}. The FRs do not experience the same broad shift in stellar mass as they are less likely to merge due to their high relative velocities within the cluster. If massive SRs grow by dry mergers during the hierarchical build-up of groups and clusters, we should expect the probability of being a SR to be closely linked to stellar mass, because galaxy mass cannot disappear and all the stellar mass that was added to a galaxy during its evolution will be easily detected observationally. We would also expect a dependence on environment, but this trend should be weaker because:
\begin{itemize}
\item Environment is more difficult to reliably measure empirically and clusters/groups are not easy to define observationally.
\item Galaxies move inside clusters, especially during merging of groups, and do not simply sit still at the centre of clusters. 
\item A massive SR could have merged with perhaps all of its nearby neighbours so that it would appear to be in a low-density environment (or even isolated) in the present day. Hence, galaxies that formed in a high density environment may sometimes be observed in lower density ones at a later stage of their evolution.
\end{itemize}
These three effects will act to wash out any environmental trend. Overall, the trends with mass and environment we observe are consistent with the overall picture described in \citetalias{cappellari2016structure}. They are also in excellent agreement with results from numerical simulations \citep{choi2018spin}.\par
In \cref{sec:comparisonwork}, we attempted to replicate the results published by three independent studies on the effect of environment on stellar angular momentum. All three studies account for the degeneracy between stellar mass and environment by investigating trends at fixed intervals of stellar mass. While there is nothing wrong with this approach, the results can be open to interpretation, considering that the errors tend to increase when the sample size is reduced. We do not suffer from this problem due to our large sample size. Furthermore, the flat selection in stellar mass is what allows us to sample a wide range in environment of over four orders of magnitude in number density. Combining the clear trends seen in Figure \ref{fig:veale}, \cref{fig:greene} and \cref{fig:brough}, we conclude that the fraction of SRs \textit{does} increase with $\Sigma_3$ at fixed mass, but that the trends are only valid if the mass interval is above $M_{\rm{crit}}$. The most robust trend is seen in the bottom left panel of \cref{fig:greene} where we show F(SR) as a function of $\Sigma_3$ for massive galaxies in the complete, volume-limited sample. The clear increase of F(SR) with increasing local density is strong evidence that environment does affect the likelihood of a galaxy being a SR. The trend is robust against our misclassification bias.\par
It is difficult to quantify to what extent each of the effects discussed in this section contribute to the discrepancy between the results presented in \cref{sec:comparisonwork} compared to each respective comparison study. For example, it is unclear whether the ``washing out" of the kT-$\Sigma$ relation in the \textit{comparison} studies is wholly responsible for their lack of reported trends in angular momentum, or if the definition of environment also plays a role (for example using $\Sigma_5$ versus $\Sigma_3$). Certainly, our larger sample size and careful catalogue construction give us confidence that the environment we obtain gives an accurate reflection of the true intrinsic environment of MaNGA galaxies and their neighbours. As environment is comparatively difficult to measure reliably compared to other quantities such as stellar mass, an apparent lack of observed trends does not necessarily rule out such a trend existing, especially if the sample size is not sufficiently large to have meaningful numbers of galaxies in each mass-environment bin.

%% file: Conclusions3.tex
\section{Conclusions}
We have presented results from the largest complete study of galaxy angular momentum and environment to date. About one third of our sample have MaNGA stellar kinematic observations, which itself represents the largest homogenous sample with stellar kinematics (\cref{fig:lambda_ellip}), and of the remaining two-thirds, only about 2\% have uncertain angular momentum classifications, meaning that the sample overall is robust.\par
As a starting point, we reproduce the kT-$\Sigma$ relation by calculating the fraction of galaxies that are classified as SRs \textit{and} are more massive than a critical stellar mass of $2\times10^{11} \textrm{ M}_{\odot}$, F(SR), in bins of $\log(\Sigma_3)$ where $\Sigma_3$ is a projected local density estimator. These galaxies are the relics of violent dry merger events experienced by their progenitors as a result of being embedded at the centres of the most massive dark matter halos at the time of their formation. Indeed, we see a clear trend where F(SR) increases with increasing local density (\cref{fig:Group_mem_sigma_3}).\par
In \citetalias{graham2019bcatalogue}, we introduced a new parameter $\Sigma_3^{\rm{rel}}$ which is a \textit{relative} density estimator. Our motivation for defining a density that is relative to the peak of the group or cluster is that the formation of core SRs is inextricably linked to the formation of the clusters themselves. The maximum density achieved in clusters increases with the cluster richness (\cref{fig:Group_mem_sigma_3}) and so the individual F(SR) profiles for each cluster are smoothed out if combined into a single kT-$\Sigma$ relation as we have done in \cref{fig:Group_mem_sigma_3}. We recalculate F(SR) as a function of $\Sigma_3^{\rm{rel}}$ and find that the massive SR distribution is skewed more towards the \textit{peak} densities rather than simply \textit{high} densities (\cref{fig:Group_mem_sigma_3_rel}). Hence, we find clear evidence that core SRs, which are the relics of dry merger events and are selected here to be above a critical stellar mass of $2\times10^{11} \textrm{ M}_{\odot}$, are strongly biased towards the dense cores of clusters and groups. This conclusion agrees with what we see by eye for individual clusters \citepalias{graham2019dclusters} and is one of the key results from this work. Because about a quarter of all SRs will be misclassified (see \citetalias{graham2019bcatalogue}), we check to see if the trend with $\Sigma_3^{\rm{rel}}$ remains after taking this into account. We find that while the trend becomes weaker, it does not disappear completely (\cref{fig:random_sr}). In total, the uncertainty in F(SR) is less than about 1.5\% due to measurement and classification errors.\par
We also investigate F(SR) as a function of distance to the centre of the cluster (\cref{fig:Group_mem_dist}) and normalised cluster-centric radius (\cref{fig:Group_mem_dist_normed}) for the complete sample and find stronger trends than those based on $\Sigma_3$ and $\Sigma_3^{\rm{rel}}$. This may suggest that the distance to the group centre is a better predictor of the abundance of massive SRs than density-based estimators.\par
As $\Sigma_3^{\rm{rel}}$ is a more fundamental parameter than $\Sigma_3$, we investigate the hypothesis that the kT-$\Sigma^{\rm{rel}}$ relation is simply a result of dynamical friction, where the most massive galaxies sink to the bottom of the potential well due to their stellar mass. In \cref{fig:sigma_rel_fixed_mass}, we calculate the fraction of SRs in four bins of stellar mass ($M \geq M_{\rm{crit}}$) and find that in all bins, F(SR) increases with increasing $\Sigma^{\rm{rel}}$. The trend deviates from flatness within the errors and so we can confidently rule out the dynamical friction hypothesis. We also studied the stellar mass distribution of massive SRs and FRs as a function of group richness (proportional to halo mass) in \cref{fig:mass-radius-dist}. If the large galaxy clusters are built up from smaller groups as is predicted by hierarchical structure formation, and if the present-day SRs were always at the centres of the ``building blocks", then we would expect to see an increase in the stellar mass of massive SRs with increasing group richness. This is exactly what we see in \cref{fig:mass-radius-dist} supporting the above hypothesis. The FRs (including spirals) do not show a similar increase which is consistent with the idea that they do not follow the dry merger channel and are unlikely to merge in the cluster environment.\par
Finally, we aim to reproduce the results of previous studies that did not find a trend of angular momentum with environment at fixed stellar mass (\cref{sec:comparisonwork}). We do find evidence for such a trend, and because we have better statistics than the previous studies did thanks to our large same size and completeness, we are confident that these trends do exist. However, we are in some cases limited by the fact that in order to investigate trends as a function of $\lambda_{R_e}$, we must restrict ourselves to the MaNGA sample which is not a volume weighted sample. We could weight the MaNGA sample by volume but this tends to amplify unusual objects at the expense of massive SRs. However, we see evidence for the same trends in the volume-weighted sample when we simply consider F(SR), which of course is related to $\lambda_{R_e}$.\par 
Our analysis has focused on the relatively nearby Universe ($z\leq0.08$) and so a natural extension to this work would be to target clusters at high redshift with the next generation of IFUs. Our hypothesis suggests that the core SRs are formed at early times and so the progenitor clusters of present day clusters should display similar morphologies to the ones in our sample. Filling in these gaps will provide an important constraint on galaxy evolution models and the theory of slow rotator formation.